\documentclass[pre,aps,twocolumn,showpacs,floatfix,superscriptaddress]{revtex4-2}
\usepackage{graphicx}
\usepackage{bm}
\usepackage{color}
\usepackage{amssymb,amsmath}
\usepackage{ulem}
\usepackage{ragged2e}
\usepackage{adjustbox}
\usepackage{multirow}
\usepackage{booktabs}
\usepackage[table]{xcolor}
\usepackage{tcolorbox}
\usepackage{tabularx}
\usepackage{array}
\usepackage{placeins}
\usepackage{colortbl}
\usepackage{hyperref}
\hypersetup{colorlinks=true, citecolor=blue, urlcolor=blue, linkcolor=blue}
\tcbuselibrary{skins}
\definecolor{Salmon}{RGB}{250,128,114}

\newcolumntype{Y}{>{\raggedleft\arraybackslash}X}

\tcbset{tab1/.style={\centering, fonttitle=\bfseries\small,fontupper=\normalsize\sffamily,
		colback=yellow!10!white,colframe=red!75!black,colbacktitle=Salmon!40!white,
		coltitle=black,center title,freelance,frame code={
			\foreach \n in {north east,north west,south east,south west}
			{\path [fill=red!75!black] (interior.\n) circle (3mm); };},}}

\tcbset{tab2/.style={enhanced,fonttitle=\bfseries,fontupper=\small\sffamily,
		colback=yellow!10!white,colframe=red!50!black,colbacktitle=Salmon!40!white,
		coltitle=black,center title}}
\begin{document}	
	\title{Frustration-Induced Collective Dynamical States in Pulse-Coupled Adaptive Winfree Networks}
	\author{R Anand}
	\email{anand19032002@gmail.com}
	\affiliation{Centre for Nonlinear Science and Engineering, Department of Physics, School of EEE, SASTRA Deemded to be University, Thanjavur, 613 401. }
	\author{V K Chandrasekar}
	\email{chandrasekar@eee.sastra.edu}
	\affiliation{Centre for Nonlinear Science and Engineering, Department of Physics, School of EEE, SASTRA Deemded to be University, Thanjavur, 613 401. }
	\author{R Suresh}
	\email{suresh@eee.sastra.edu}
	\affiliation{Centre for Nonlinear Science and Engineering, Department of Physics, School of EEE, SASTRA Deemded to be University, Thanjavur, 613 401. }

	\begin{abstract}
We investigate collective dynamics in a pulse-coupled adaptive Winfree network under the influence of a frustration (phase-lag) parameter. The coupling strengths coevolve according to a Hebbian adaptation rule and self-organize to support a wide variety of collective states. We observe frequency-clustered states, entrainment, bump states, bump--frequency cluster states, antipodal and multi-antipodal cluster states, chimera states, and incoherent dynamics. Notably, we report for the first time the spontaneous emergence of entrainment, bump, and bump--frequency cluster states in an adaptive network {\it without} any external forcing.  To systematically characterize these regimes, we introduce three complementary measures of incoherence based on (i) time-averaged frequencies, (ii) instantaneous phases, and (iii) mean frequencies per bin. These measures enable the construction of one- and two-parameter phase diagrams that clearly delineate transitions between distinct dynamical states. Furthermore, we analytically derive the stability condition for the frequency-entrained state, which shows excellent agreement with numerical simulations. Our results highlight the crucial role of frustration-mediated plasticity in shaping rich self-organized dynamics in pulse-coupled adaptive networks.
\end{abstract}
	
	\maketitle
\section{Introduction} \label{intro}
\par The self-organization of dynamical phenomena across multiple scales is a fundamental feature of nonequilibrium systems and is widely observed in plasma physics, turbulence, and neuroscience \cite{bak1988self, hinrichsen2000non, hasegawa1987self}. One of the simplest manifestations of such multiscale hierarchical organization is the spontaneous emergence of coordinated collective patterns from underlying microscopic disorder, a process closely analogous to phase transitions in equilibrium systems \cite{kaneko1990globally}. A prominent and extensively studied example of this emergent collective behavior is macroscopic synchronization in ensembles of coupled oscillators. Despite inevitable differences in intrinsic frequencies, interactions can induce spontaneous phase locking, enabling a subset of oscillators to oscillate collectively at the same frequency \cite{strogatz2000kuramoto,pikovsky2001universal}. Striking examples of collective synchronization are widespread in nature, ranging from the coordinated murmurations of starling flocks \cite{bialek2012statistical,das2024flocking} to the precisely timed synchronous flashing of fireflies \cite{buck1966biology,buck1988synchronous}, phenomena that have attracted long-standing scientific interest.

In 1967, Arthur T.~Winfree introduced a seminal mathematical model that laid the foundation for understanding synchronization in populations of biological oscillators \cite{winfree1967biological}. This framework demonstrated how weakly coupled oscillators, each possessing its own intrinsic rhythm, can achieve spontaneous collective synchrony through brief, pulse-like interactions that emulate impulsive stimulation in natural systems such as pacemaker cells and flashing fireflies. Subsequently, Kuramoto proposed a simplified model in which interactions are represented by sinusoidal functions of phase differences, greatly enhancing analytical tractability at the expense of some biological realism \cite{kuramoto2005self}. For a particularly tractable class of pulse-coupled oscillators, Ariaratnam {\it et al.} carried out a detailed bifurcation analysis of the Winfree model, revealing rich dynamical behavior through comprehensive phase diagrams \cite{ariaratnam2001phase}.

The Kuramoto model owes its status as a cornerstone of synchronization research, including contemporary studies on adaptive networks, to its analytical feasibility \cite{seliger2002plasticity,ren2007adaptive,aoki2009co,aoki2011self,abbott2000synaptic}. More recently, attention has shifted toward adaptive networks, in which coupling strengths or network topology coevolve with the dynamics of the nodes \cite{taylor2010spontaneous,horstmeyer2020adaptive,markram1997regulation,abbott2000synaptic,jain2001model,kuehn2015multiple}. Such adaptation mechanisms introduce feedback between structure and dynamics, promoting self-organized synchronization and a wide range of complex collective phenomena. When adaptation evolves on a timescale much slower than the intrinsic oscillator dynamics, singular perturbation theory can be employed to elucidate how long-term plasticity sculpts emergent collective states \cite{kuehn2015multiple}.

A growing body of work has explored self-organized dynamical patterns in adaptive networks. The slow nature of adaptation is essential for maintaining such multi-cluster configurations, where short-lived structures exert a strong influence on the eventual collective behavior \cite{berner2019hierarchical}. Adaptive networks are capable of sustaining partially synchronized or hierarchically organized states, including splay, antipodal, and double-antipodal configurations, through a subtle interplay between evolving couplings and network topology \cite{berner2020birth}. Similar mechanisms have been identified across diverse adaptive systems, ranging from Morris--Lecar neuron models \cite{popovych2015spacing} to phase oscillators with time-varying couplings \cite{kasatkin2017self,kasatkin2018synchronization,kasatkin2018effect}. Recent studies have further demonstrated heterogeneous nucleation effects \cite{fialkowski2023heterogeneous} and frequency-dependent synchronization transitions in multilayer adaptive networks \cite{yadav2025heterogeneous}. Augustsson and Martens \cite{augustsson2024co} investigated coevolutionary dynamics in a minimal network of two adaptively coupled Theta neurons, revealing bistability, mode locking, and chaos induced by increasing adaptivity.

Despite its importance, the interplay between adaptation and frustration has received comparatively limited attention in pulse-coupled adaptive network studies. In this context, the Winfree model offers particular advantages. Adaptation rules in biological systems are often motivated by synaptic plasticity mechanisms, especially the dependence on spike timing between pre- and postsynaptic neurons \cite{gerstner1998neuronal,caporale2008spike,bi2001synaptic,meisel2009adaptive,lucken2016noise}. The most prominent examples include Hebbian learning and spike-timing-dependent plasticity (STDP). Encapsulated in the adage ``neurons that fire together wire together'' \cite{donald1949organization,lowel1992selection,zhang1998critical}, Hebbian rules promote the strengthening of couplings for coherent activity and their weakening for out-of-phase interactions. These principles are central to theories of memory formation and neuromorphic engineering \cite{acebron2005kuramoto,pickett2013scalable,hansen2017double,birkoben2020slow}. A deeper understanding of how plasticity interacts with phase lag is therefore crucial for explaining complex behavior in coevolving dynamical systems \cite{timms2014synchronization,thamizharasan2025dynamics,augustsson2024co}.

The introduction of a phase lag in the Winfree phase-response curve (PRC) determines whether coupling advances or delays oscillator phases, thereby regulating collective synchrony. Laing {\it et al.} \cite{laing2021dynamics} demonstrated that the phase-lag parameter alters the nature and location of Hopf and SNIC bifurcations, significantly affecting the stability of synchronized states in structured networks. Consequently, phase lag emerges as a key control parameter governing the interplay between network structure and macroscopic oscillatory dynamics.

Previous studies have shown that Hebbian adaptation can generate diverse dynamical states in systems of identical phase oscillators subjected to external forcing \cite{thamizharasan2024stimulus}. In the absence of forcing, however, the reported states were limited to two-cluster, multi-antipodal cluster, splay cluster, and splay chimera configurations. Under external forcing, additional states such as synchronization, forced entrainment, chimera, and bump states were observed. Furthermore, bump--frequency cluster states have been reported in adaptive Kuramoto models driven by external forcing under hard-bounded Hebbian adaptation \cite{thamizharasan2024hebbian}.

In this study, we report the first observation of entrainment (ENT), bump (BS), and bump--frequency cluster (BFC) states emerging spontaneously in adaptive dynamical networks without any external forcing. In the ENT state, all oscillators phase lock while maintaining zero time-averaged frequency. The BFC state consists of two coexisting clusters: one exhibiting fully developed spiking oscillations at a well-defined frequency and the other displaying small-amplitude, subthreshold (bump-like) oscillations driven by the spiking cluster. In contrast, the BS state is characterized by an inactive coherent domain of quiescent oscillators coexisting with an active incoherent domain exhibiting irregular, small-amplitude oscillations. Notably, these states arise in a pulse-coupled adaptive Winfree neuron model driven solely by variations in the frustration (phase-lag) parameter. In addition, the network supports frequency-clustered, antipodal, multi-antipodal, chimera, and incoherent states.

In this work, we incorporate a Hebbian adaptation rule into the Winfree model of pulse-coupled oscillators, a biologically realistic framework that accounts for finite-duration pulsatile interactions, and explore the resulting frustration-induced collective dynamics. To characterize these behaviors, we introduce three independent measures of incoherence based on time-averaged frequencies, instantaneous phases, and mean frequencies per bin, and construct one- and two-parameter phase diagrams that reveal transitions between distinct dynamical regimes. We also provide an analytical stability analysis of the frequency-entrained state, demonstrating strong agreement with numerical simulations.

The remainder of the paper is organized as follows. Section~\ref{model} presents the model formulation of the adaptive network. In Sec.~\ref{si}, we describe the numerical methods and incoherence measures used to characterize collective states. Section~\ref{result} discusses the frustration-induced dynamical states and their organization in one- and two-parameter phase diagrams. In Sec.~\ref{Analytical}, we present an analytical stability analysis of the frequency-entrained state. Finally, Sec.~\ref{conclusion} summarizes the main findings and their implications.
	
\section{Model} \label{model}
\par We consider a pulse-coupled, identical Kuramoto--Winfree ensemble consisting of a large number ($N \gg 1$) of globally coupled phase oscillators \cite{gallego2017synchronization,ariaratnam2001phase,ermentrout1991adaptive,quinn2007singular}. The interaction strengths between oscillators evolve dynamically according to a Hebbian learning rule. The evolution of the individual phases $\theta_i$ ($i=1,\dots,N$) is governed by the following system of coupled ordinary differential equations:
\begin{eqnarray}
	\label{phase}
	\dot{\theta}_i &=& \omega + Q(\theta_i+\alpha)\frac{\sigma}{N}\sum_{\substack{j=1\\ j\neq i}}^{N} k_{ij} P(\theta_j).
\end{eqnarray}

Each oscillator receives the same global input through the mean field $S_i = N^{-1}\sum_{\substack{j=1\\ j\neq i}}^{N} P(\theta_j)$, where $P(\theta)$ is a pulse-like influence function representing the finite-duration interactions among oscillators. The phase response curve (PRC) $Q(\theta)$ determines how an oscillator responds to the mean-field signal generated by the population. Both $P(\theta)$ and $Q(\theta)$ are $2\pi$-periodic functions, defined on either $[0,2\pi)$ or $[-\pi,\pi)$. The parameter $\alpha$ denotes the phase lag (or frustration parameter). Throughout this work, the natural frequency is fixed at $\omega=1$, and the global coupling strength is set to $\sigma=1$.

The dynamical coupling weights $k_{ij}$ quantify the influence of oscillator $j$ on oscillator $i$. Their evolution is governed by a Hebbian learning rule of the form
\begin{eqnarray}
	\label{coupling}
	\dot{k}_{ij} = \epsilon\bigl[\cos(\theta_i-\theta_j)-k_{ij}\bigr],
\end{eqnarray}
where $\epsilon \ll 1$ is a small timescale parameter that enforces a clear separation between the fast phase dynamics and the slow adaptation of the coupling strengths.

\subsection{Pulse shape} \label{pulse}
The influence function $P(\theta)$, which determines the shape of the pulses emitted by each oscillator and appears explicitly in the mean-field term, is chosen to be of the Ariaratnam--Strogatz (AS) type \cite{ariaratnam2001phase,gallego2017synchronization}. This functional form is well suited for modeling sharp, pulse-like interactions commonly encountered in biological and physical systems, including networks of flashing fireflies and neuronal populations. The AS pulse function is defined as
\begin{eqnarray}
	\label{AS}
	P(\theta) = a_n\bigl(1+\cos\theta\bigr)^n,
\end{eqnarray}
where the normalization constant is $a_n = 2^n (n!)^2/(2n)!$. For the specific choice $n=1$, one obtains $a_1=1$, yielding
\begin{eqnarray}
	\label{AS1}
	P(\theta) = 1+\cos\theta.
\end{eqnarray}

This pulse function possesses several properties that highlight its pulsatile nature: it is unimodal and symmetric about $\theta=0$, attains its maximum at $\theta=0$, vanishes at $\theta=\pm\pi$, and is normalized such that $\int_{-\pi}^{\pi}P(\theta)\,d\theta = 2\pi$, ensuring unit mean influence over one oscillation cycle (see Fig.~\ref{plse}).

\subsection{Phase response curve (PRC)}
\begin{figure}[!ht]
		\centering
		\includegraphics[width=0.5\textwidth]{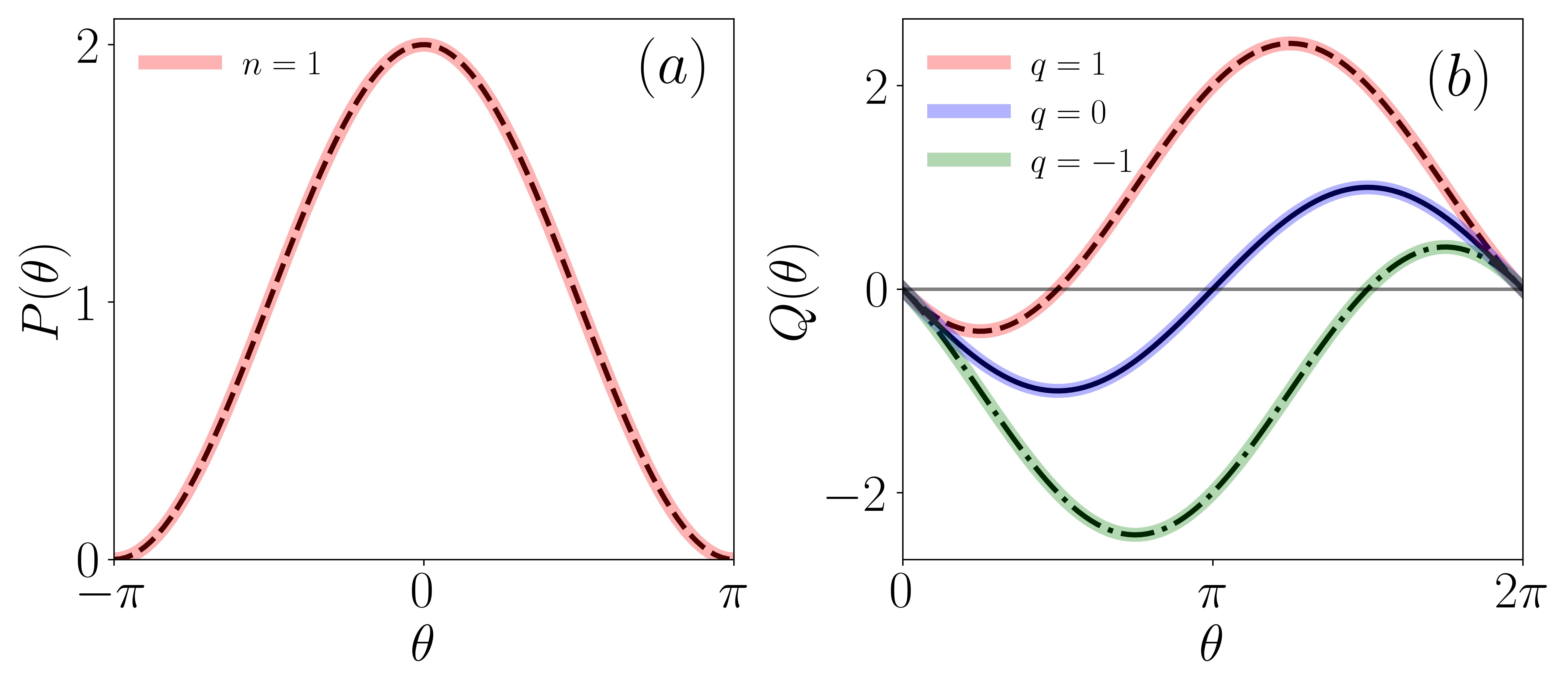}
		\caption{(a) The influence function $P(\theta)$ for $n=1$. (b) Phase response curves $Q(\theta)$ for different values of $q = 1$, $0$, and $-1$.}
		\label{plse}
	\end{figure}

The phase response curve $Q(\theta)$ characterizes the sensitivity of an oscillator to incoming pulses as a function of its phase. In this study, we adopt a sinusoidal PRC that vanishes at the instant of pulse emission, $Q(0)=0$, representing a refractory period during which the oscillator is temporarily insensitive to external perturbations immediately after firing. The PRC is chosen as
\begin{eqnarray}
	\label{PRC}
	Q(\theta) = q\bigl(1-\cos\theta\bigr)-\sin\theta,
\end{eqnarray}
where $q$ is an offset parameter controlling the balance between phase advances and delays. For $q>0$, pulse interactions predominantly induce phase advances, whereas for $q<0$ they primarily generate phase delays, as illustrated in Fig.~\ref{plse}.

To incorporate frustration or phase-lag effects \cite{laing2021dynamics}, we introduce a phase shift $\alpha$ into the PRC, such that Eq.~\eqref{PRC} becomes
\begin{eqnarray}
	\label{PRC1}
	Q(\theta+\alpha) = q\bigl[1-\cos(\theta+\alpha)\bigr]-\sin(\theta+\alpha),
\end{eqnarray}
with $\alpha\in[0,\pi/2)$. Within this range, the resulting PRC is biphasic, exhibiting both positive and negative regions, and therefore effectively captures Type-II neuronal excitability.
\section{Methods and measures} \label{si}
\par We numerically integrate the system of pulse-coupled oscillators with adaptive coupling, defined by Eqs.~\eqref{phase} and \eqref{coupling}, using a fourth-order Runge--Kutta scheme with a fixed time step $\Delta t = 0.01$. Unless otherwise stated, simulations are performed with $N=100$ oscillators. The parameters are fixed at $\omega=1$, $\sigma=1$, and $\epsilon=0.01$. Initial phases are drawn randomly from a uniform distribution on $[0,2\pi)$, while the initial coupling strengths are random uniformly distributed in the interval $[-1,1]$. Depending on the value of the frustration parameter $\alpha$, the adaptive network exhibits a variety of collective dynamical states.

To systematically characterize these states, we employ three complementary measures of incoherence \cite{gopal2014observation}: the frequency-based strength of incoherence ($S$), the instantaneous-phase-based strength ($S_\sigma$), and the mean-frequency-based strength of incoherence ($S_\omega$). For this purpose, the oscillator population is partitioned into $M$ bins, each containing $n=N/M$ oscillators. Throughout this study, we use $M=20$ bins with $n=5$ oscillators per bin.

\subsection{Frequency-based strength of incoherence}
The local standard deviation of the time-averaged frequencies within the $m$th bin is defined as
\begin{equation}
	\sigma_m = \sqrt{\frac{1}{n}\sum_{j=n(m-1)+1}^{mn}\left(\omega_j-\bar{\omega}_m\right)^2}, 
	\qquad m=1,2,\dots,M,
	\label{eq:sig_m}
\end{equation}
where
\[
\bar{\omega}_m=\frac{1}{n}\sum_{j=n(m-1)+1}^{mn}\omega_j
\]
is the mean frequency of the $m$th bin, and $\omega_j$ denotes the time-averaged frequency of the $j$th oscillator.

The frequency-based strength of incoherence is then defined as
\begin{equation}
	S = 1-\frac{1}{M}\sum_{m=1}^{M}s_m, 
	\qquad 
	s_m=\Theta(\zeta-\sigma_m),
	\label{eq:S}
\end{equation}
where $\Theta$ is the Heaviside step function and $\zeta=0.005$ is a fixed threshold used throughout this work.

For frequency-entrained states, including antipodal and entrainment states, $S=0$. For frequency-clustered states, $S=N_{\mathrm{FC}}/M\neq 0$, where $N_{\mathrm{FC}}$ denotes the number of frequency clusters. Chimera and bump states satisfy $0<S<1$, while frequency-cluster, multi-antipodal, and bump--frequency clustertypically yield $0<S\ll 1$. In the fully incoherent state, $S=1$.

\subsection{Phase-based strength of incoherence}
To quantify local phase coherence, we compute the standard deviation of the instantaneous phases within each bin,
\begin{equation}
	\hat{\sigma}_m = \sqrt{\frac{1}{n}\sum_{j=n(m-1)+1}^{mn}\left(\theta_j-\bar{\theta}_m\right)^2},
	\qquad m=1,2,\dots,M,
	\label{eq:smb}
\end{equation}
where
\[
\bar{\theta}_m=\frac{1}{n}\sum_{j=n(m-1)+1}^{mn}\theta_j
\]
is the mean instantaneous phase of the $m$th bin and $\phi_j$ denotes the instantaneous phase of the $j$th oscillator.

The phase-based strength of incoherence is defined as
\begin{equation}
	S_\sigma = 1-\frac{1}{M}\sum_{m=1}^{M}\bar{s}_m,
	\qquad 
	\bar{s}_m=\Theta(\delta-\hat{\sigma}_m),
	\label{eq:Ssig}
\end{equation}
where the threshold is fixed at $\delta=0.05$. For the incoherent state, $S_\sigma=1$. Antipodal, frequency-cluster, and bump--frequency cluster states satisfy $0<S_\sigma\ll<1$. Multi-antipodal, bump, and chimera states satisfy $0<S_\sigma<1$, whereas the entrainment state is characterized by $S_\sigma=0$.

\subsection{Mean-frequency-based strength of incoherence}
The mean-frequency-based strength of incoherence is defined as
\begin{equation}
	S_\omega = 1-\frac{1}{M}\sum_{m=1}^{M}\hat{s}_m,
	\qquad 
	\hat{s}_m=\Theta(\xi-\bar{\omega}_m),
	\label{eq:Somg}
\end{equation}
where $\xi=0.005$. The entrainment state is uniquely characterized by $S_\omega=0$. For bump and bump--frequency cluster states, $0<S_\omega<1$, whereas all other dynamical states considered in this study yield $S_\omega=1$.
	\begin{table}[!ht]
	\centering
	\renewcommand{\arraystretch}{1.8}
	\setlength{\tabcolsep}{3pt}
	\begin{tabular}{cccc}
		\hline\hline
		Dynamical state & $S$ & $S_\sigma$ & $S_\omega$ \\ \hline
		Anti-podal & $S = 0$ & $0 < S_\sigma \ll 1$ & $S_\omega = 1$ \\
		Frequency cluster & $0 < S \ll 1$ & $0 < S_\sigma \ll 1$ & $S_\omega = 1$ \\
		Multi-antipodal & $0 < S \ll 1$ & $0 < S_\sigma < 1$ & $S_\omega = 1$ \\
		Chimera & $0 < S < 1$ & $0 < S_\sigma < 1$ & $S_\omega = 1$ \\
		Bump-frequency cluster & $0 < S \ll 1$ & $0 < S_\sigma \ll 1$ & $0 < S_\omega < 1$ \\
		Bump state & $0 < S < 1$ & $0 < S_\sigma < 1$ & $0 < S_\omega < 1$ \\
		Entrainment & $ S = 0$ & $S_\sigma = 0$ & $S_\omega = 0$ \\
		Incoherent & $S = 1$ & $S_\sigma = 1$ & $S_\omega = 1$ \\ \hline\hline
	\end{tabular}
	\caption{Characterization of the observed dynamical states using the three incoherence measures ($S$, $S_\sigma$, and $S_\omega$).}
	\label{tab1}
\end{table}

The characteristic ranges of the three incoherence measures for the different dynamical states are summarized in Table~\ref{tab1}.	
	\begin{figure}[!ht]
		\centering
		\includegraphics[width=0.5\textwidth]{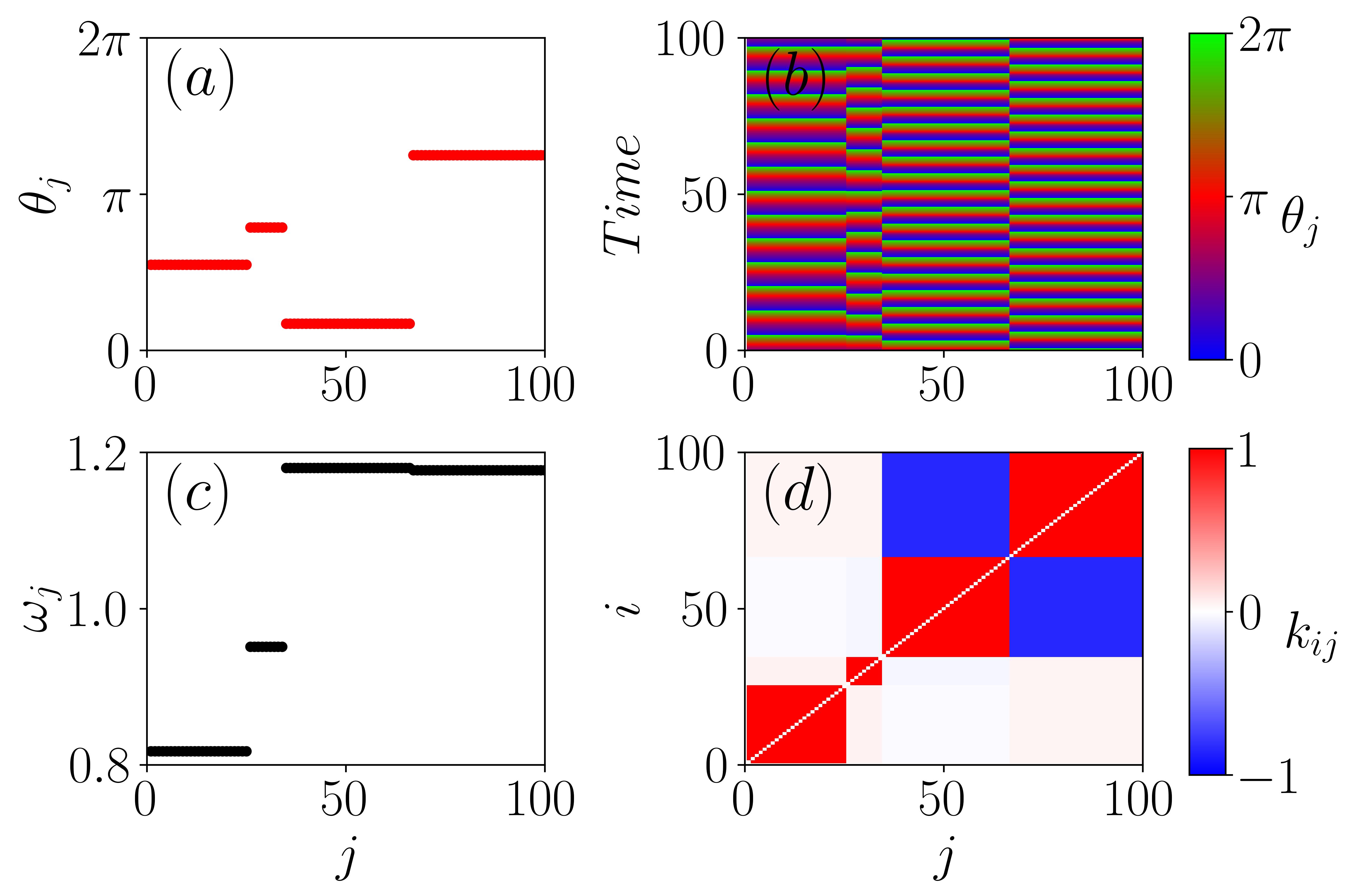}
		\caption{Frequency cluster (FC) state for $\alpha = 0$ and $q = -1$: (a) snapshot of instantaneous phases, (b) space--time evolution of the oscillators, (c) time-averaged frequency distribution, and (d) coupling matrix $k_{ij}$. Other parameters are fixed at $\omega = 1$, $\sigma = 1$, and $\epsilon = 0.01$..}
		\label{FC1}
	\end{figure}
\section{Results and Discussion} \label{result}

\subsection{Negative PRC offset $(q<0)$}
\par In this subsection, we analyze the collective dynamical states exhibited by the pulse-coupled adaptive network for a negative PRC offset ($q<0$) as the frustration parameter (phase lag) $\alpha$ is varied. The observed states are identified through a combined analysis of instantaneous phase snapshots, time-averaged frequency distributions, space--time plots, and coupling matrices. Owing to the slow adaptation rate ($\epsilon \ll 1$), all simulations are performed after discarding transients, and statistical stationarity is ensured over the time interval $t\in(26400,26500)$.

We first consider the case $\alpha=0$. For negative PRC offset, the system settles into a frequency cluster (FC) state, in which oscillators split into multiple groups oscillating at distinct frequencies \cite{thamizharasan2022exotic}. This clustering is clearly visible in the instantaneous phase snapshot [Fig.~\ref{FC1}(a)] and is further confirmed by the discrete bands in the time-averaged frequency distribution shown in Fig.~\ref{FC1}(c). The corresponding space--time plot and coupling matrix, displayed in Figs.~\ref{FC1} (b) and ~\ref{FC1}(d), exhibits a block-like structure consistent with frequency-based clustering.
\begin{figure}[!ht]
	\centering
	\includegraphics[width=0.5\textwidth]{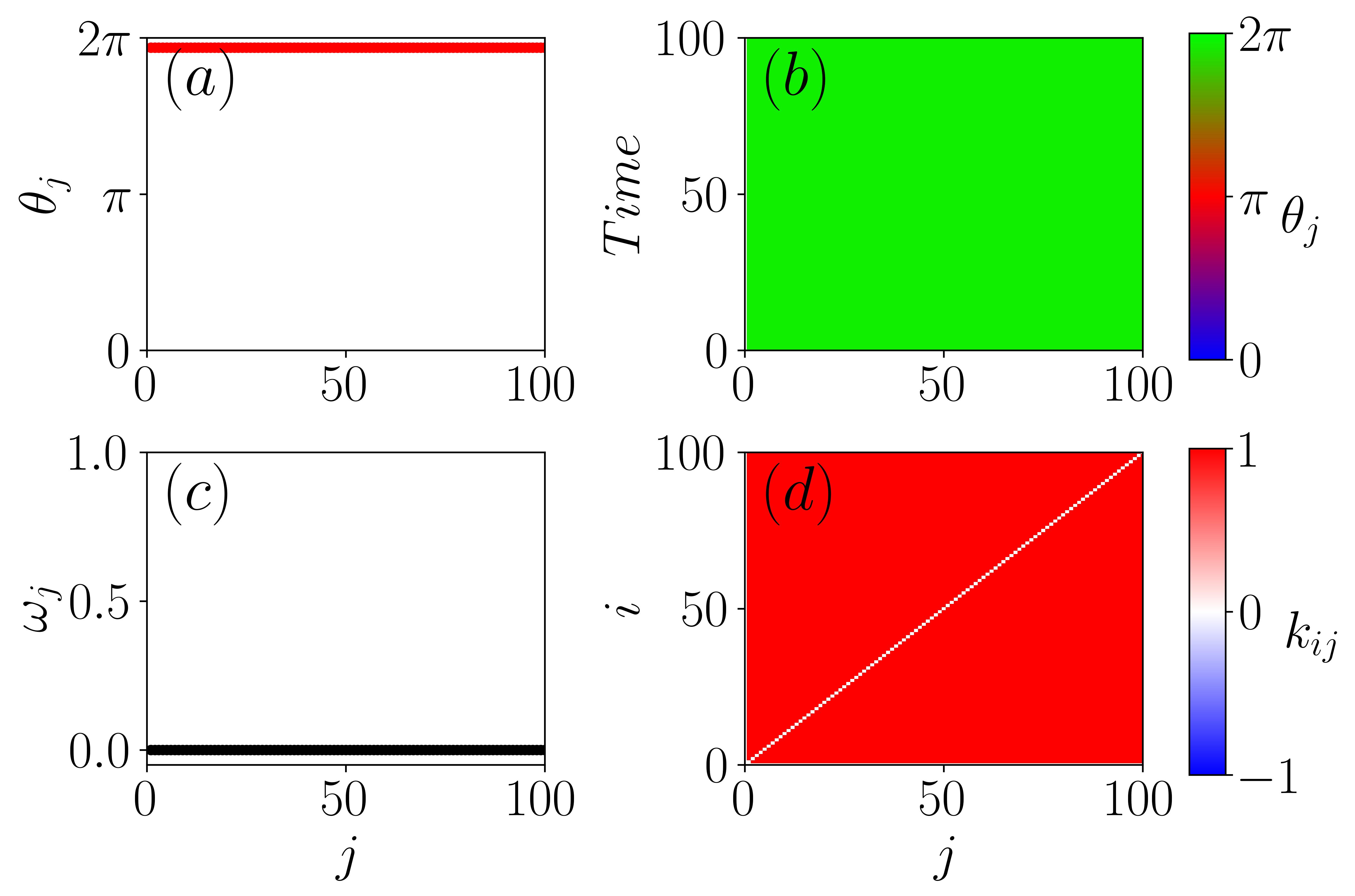}
	\caption{Entrainment (ENT) state for $\alpha = 0.15\pi$ and $q = -1$: (a) snapshot of instantaneous phases, (b) space--time evolution of the oscillators, (c) time-averaged frequency distribution, and (d) coupling matrix $k_{ij}$. All other parameters are the same as in Fig.~\ref{FC1}.}
	\label{ENT}
\end{figure}

Upon increasing the phase lag to $\alpha=0.15\pi$, the system undergoes a transition to the entrainment (ENT) state. In this regime, all oscillators become phase locked and exhibit zero time-averaged frequency, indicating complete frequency entrainment \cite{thamizharasan2024hebbian,thamizharasan2024stimulus}. The instantaneous phase snapshot and time-averaged frequency distribution for the ENT state are shown in Figs.~\ref{ENT}(a) and \ref{ENT}(c), respectively, while the corresponding space--time plot in Fig.~\ref{ENT}(b) confirms the stationary nature of the entrained dynamics. The coupling matrix in Fig.~\ref{ENT}(d) reveals uniform positive couplings ($k_{ij}=1$), reflecting the fully synchronized configuration.

\begin{figure}[!ht]
	\centering
	\includegraphics[width=0.5\textwidth]{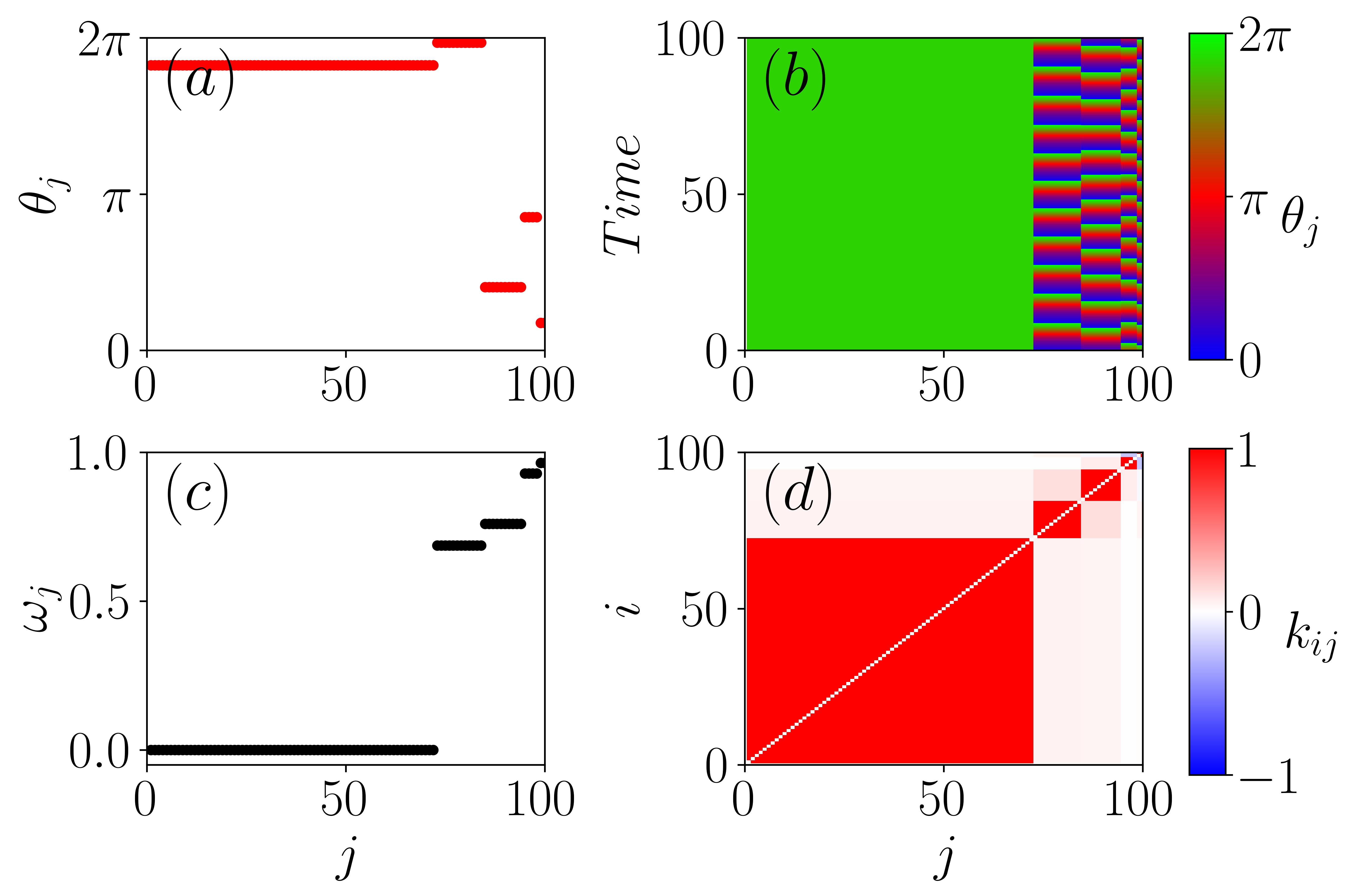}
	\caption{Bump–frequency cluster (BFC) state for $\alpha = 0.3\pi$ and $q = -1$: (a) snapshot of instantaneous phases, (b) space--time evolution of the oscillators, (c) time-averaged frequency distribution, and (d) coupling matrix $k_{ij}$. All other parameters are the same as in Fig.~\ref{FC1}.}
	\label{BFC}
\end{figure}

As the phase lag is increased further to $\alpha=0.3\pi$, the system transitions into a bump--frequency cluster (BFC) state. This hybrid state is characterized by the coexistence of two distinct dynamical groups: one cluster exhibiting fully developed spiking oscillations at a well-defined frequency and another cluster displaying small-amplitude, subthreshold (bump-like) oscillations driven by the spiking population \cite{thamizharasan2022exotic,thamizharasan2024hebbian}. The space--time plot in Fig.~\ref{BFC}(f) clearly illustrates this coexistence. The instantaneous phase snapshot and time-averaged frequency distribution are shown in Figs.~\ref{BFC}(a) and \ref{BFC}(c), respectively, while the coupling matrix in Fig.~\ref{BFC}(d) reflects the heterogeneous interaction structure.

\begin{figure}[!ht]
	\centering
	\includegraphics[width=0.5\textwidth]{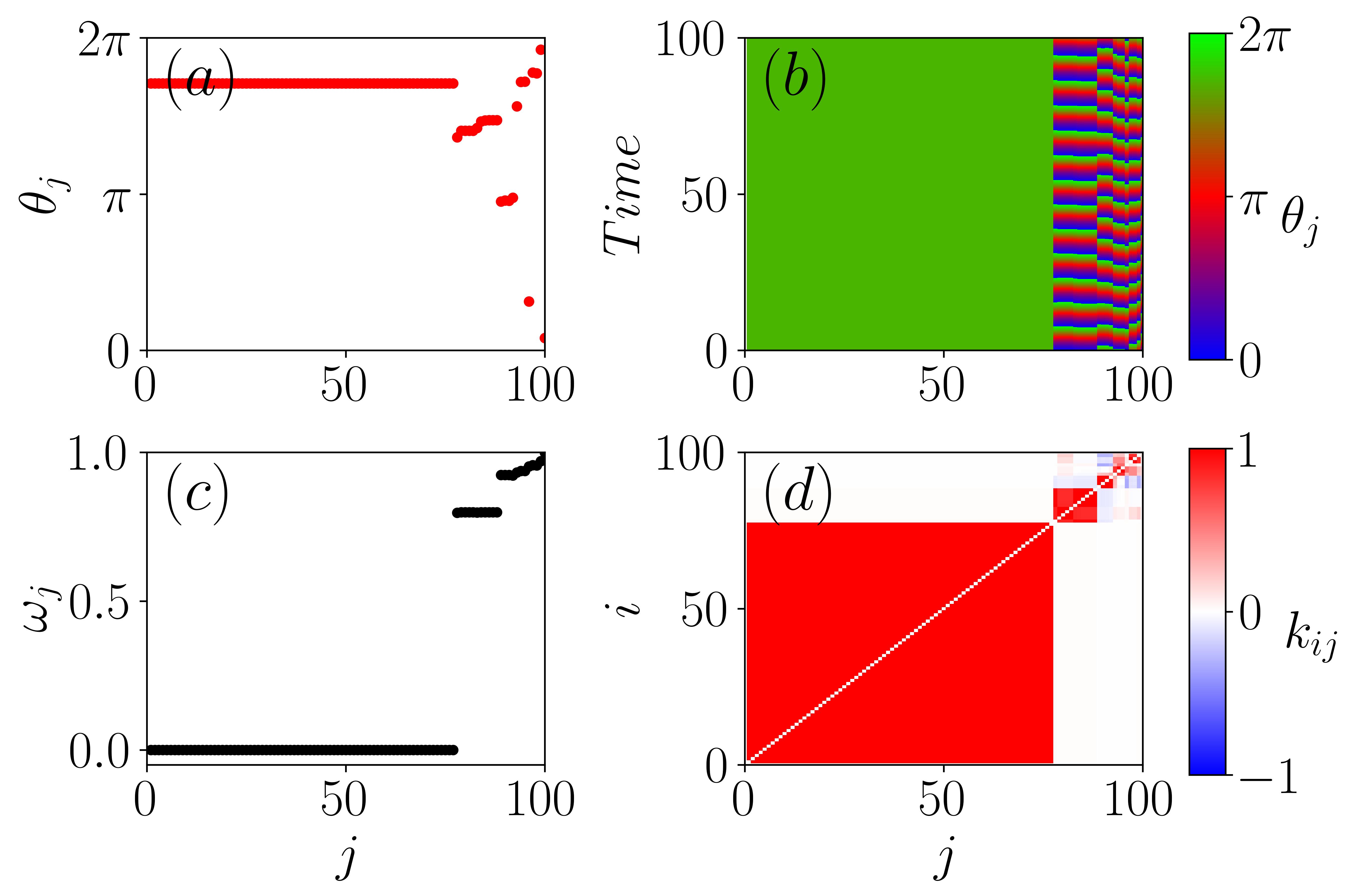}
	\caption{Bump state (BS) for $\alpha = 0.5\pi$ and $q = -1$: (a) snapshot of instantaneous phases, (b) space--time evolution of the oscillators, (c) time-averaged frequency distribution, and (d) coupling matrix $k_{ij}$. All other parameters are the same as in Fig.~\ref{FC1}.}
	\label{BS}
\end{figure}

For larger phase lag values, specifically $\alpha=0.5\pi$, the system enters the bump state (BS). This state is characterized by the coexistence of an inactive coherent domain, where oscillators remain locked in a quiescent subthreshold regime, and an active incoherent domain, where oscillators exhibit irregular, small-amplitude oscillations \cite{thamizharasan2024stimulus}. These features are evident in the space--time plot shown in Fig.~\ref{BS}. The corresponding instantaneous phase snapshot, time-averaged frequency distribution, and coupling matrix are presented in Figs.~\ref{BS}(a), \ref{BS}(c), and \ref{BS}(d), respectively.

\begin{figure}[!ht]
	\centering
	\includegraphics[width=0.5\textwidth]{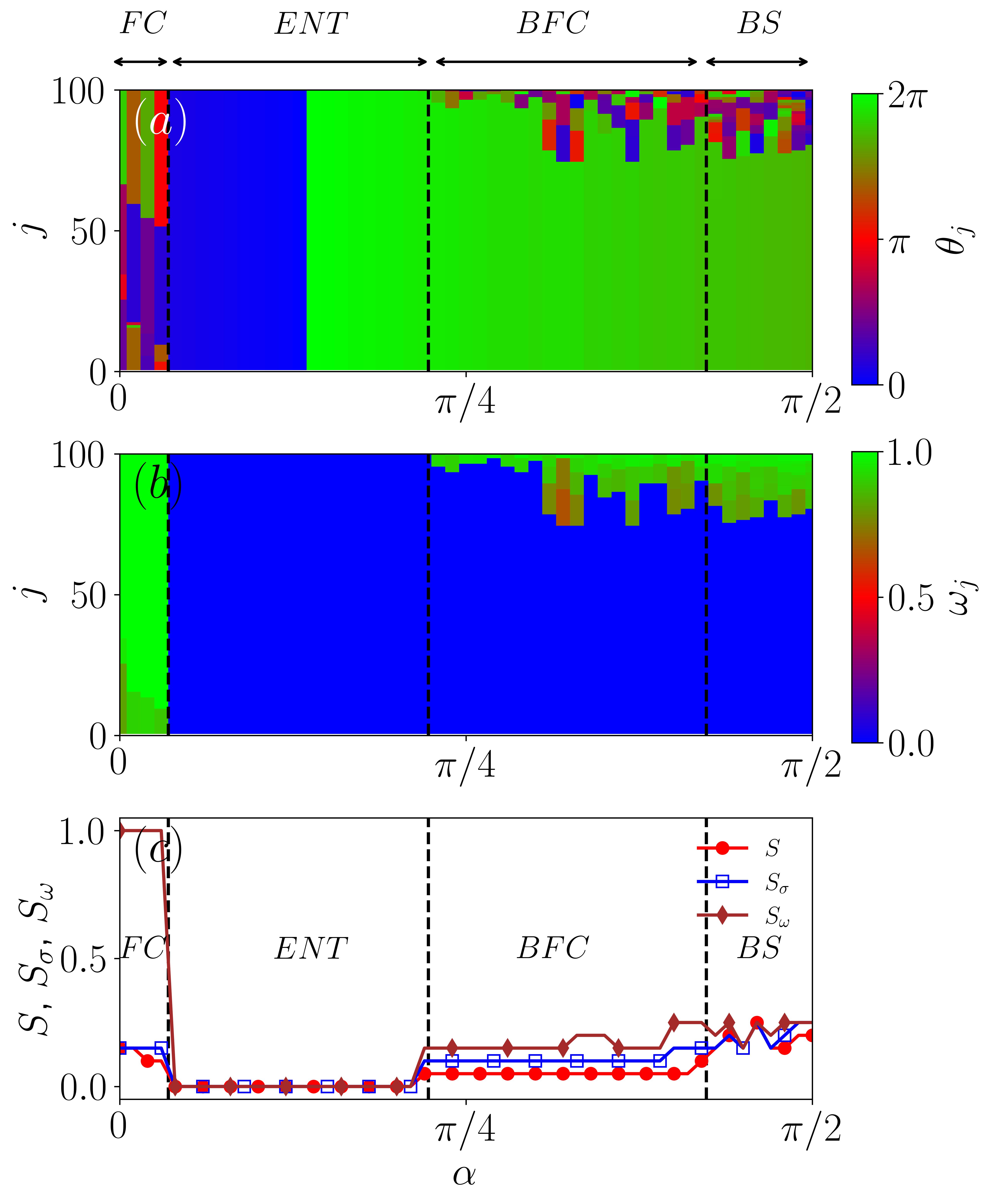}
	\caption{Dynamical transitions as a function of the phase lag $\alpha$ for $q=-1$: (a) instantaneous phases of the oscillators (color-coded by phase value), (b) time-averaged oscillator frequencies, and (c) variation of the three incoherence measures ($S$, $S_\sigma$, and $S_\omega$) with $\alpha$. All other parameters are the same as in Fig.~\ref{FC1}.}
	\label{1p1}
\end{figure}

The sequence of dynamical transitions observed for $q=-1$ is summarized in the one-parameter bifurcation diagram shown in Fig.~\ref{1p1} as a function of the phase lag $\alpha$. Figures~\ref{1p1}(a) and \ref{1p1}(b) display the instantaneous phases and time-averaged frequencies, respectively, while Fig.~\ref{1p1}(c) shows the corresponding incoherence measures ($S$, $S_\sigma$, and $S_\omega$). Specifically, the FC state appears for $\alpha\in(0,0.05\pi)$, followed by the ENT state for $\alpha\in(0.05\pi,0.23\pi)$. The BFC state emerges in the interval $\alpha\in(0.23\pi,0.4\pi)$, and the BS state is observed for $\alpha>0.42\pi$. These transitions are quantitatively supported by the distinct signatures of the incoherence measures in Fig.~\ref{1p1}(c).
\subsection{Balanced PRC $(q=0)$}
\par We now examine the collective dynamics for a balanced PRC ($q=0$), where excitatory and inhibitory contributions cancel on average and the phase lag $\alpha$ acts as the primary control parameter.
\begin{figure}[!ht]
	\centering
	\includegraphics[width=0.5\textwidth]{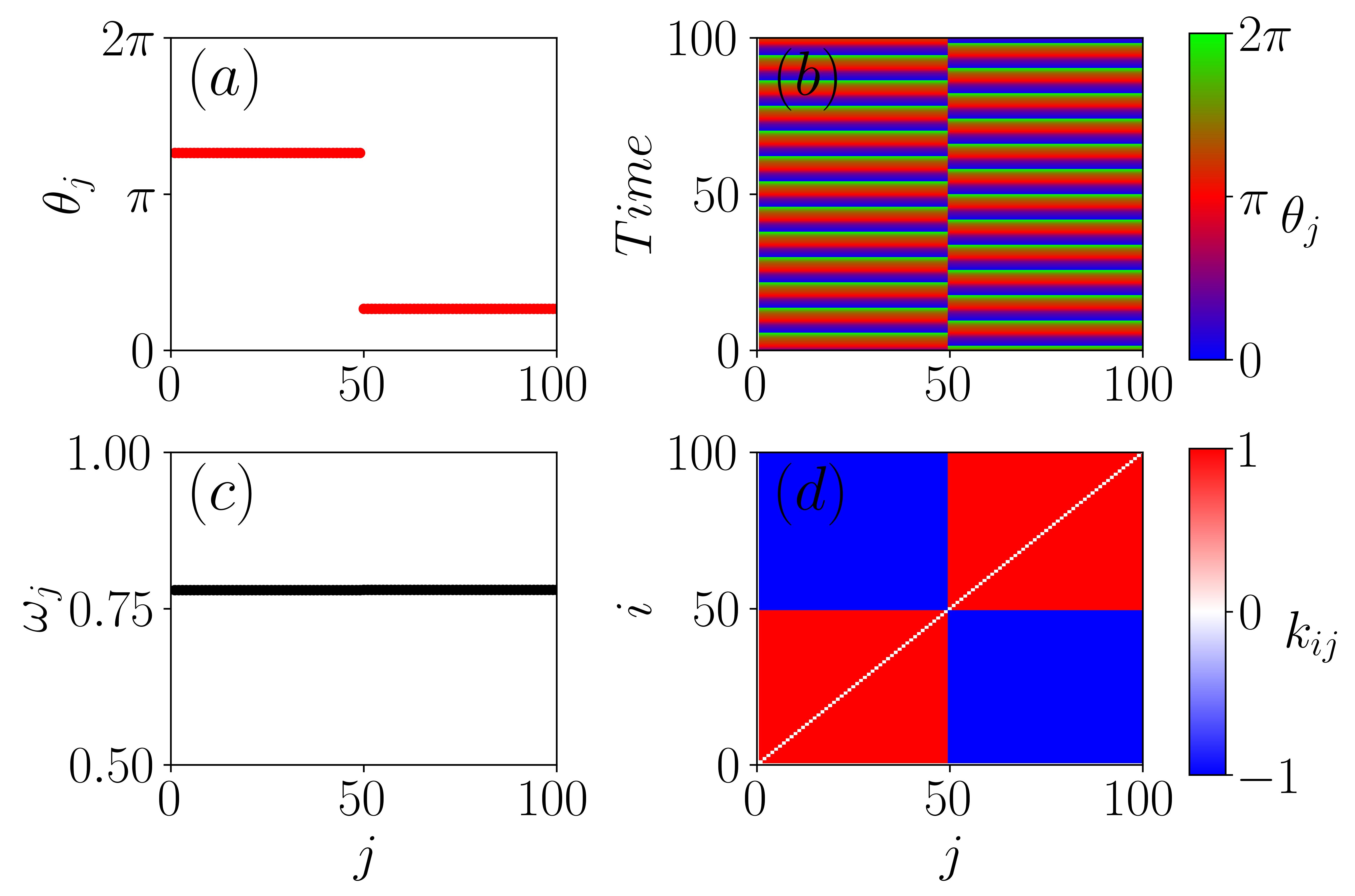}
	\caption{Antipodal (AP) state for $\alpha = 0$ and $q = 0$: (a) snapshot of instantaneous phases, (b) space--time evolution of the oscillators, (c) time-averaged frequency distribution, and (d) coupling matrix $k_{ij}$. All other parameters are the same as in Fig.~\ref{FC1}.}
	\label{AP}
\end{figure}
\begin{figure}[!ht]
	\centering
	\includegraphics[width=0.5\textwidth]{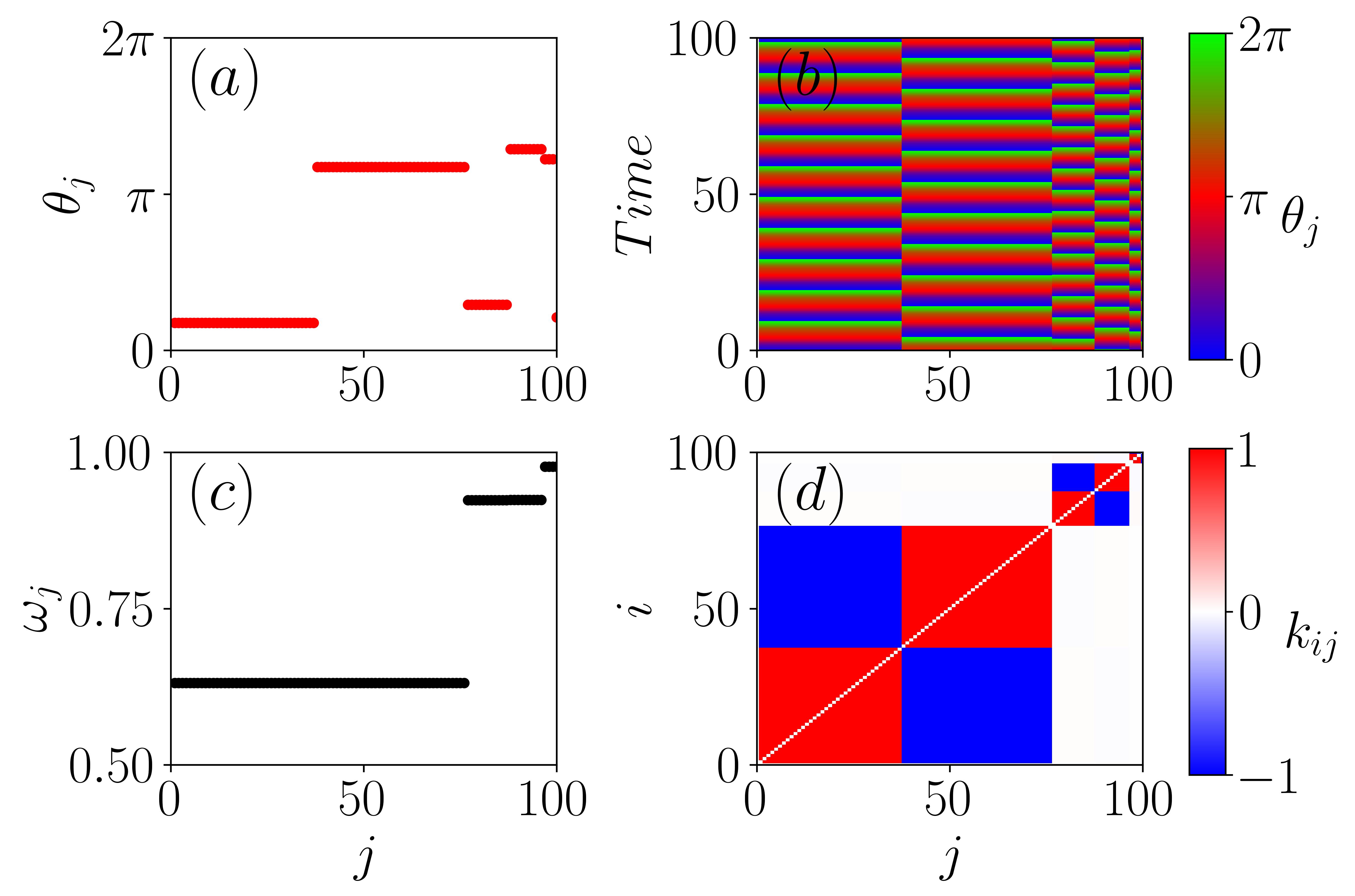}
	\caption{Multi-antipodal cluster (MAC) state for $\alpha = 0.25\pi$ and $q = 0$: (a) snapshot of instantaneous phases, (b) space--time evolution of the oscillators, (c) time-averaged frequency distribution, and (d) coupling matrix $k_{ij}$. All other parameters are the same as in Fig.~\ref{FC1}.}
	\label{MAC}
\end{figure}

For $\alpha=0$, the system organizes into the antipodal (AP) state, consisting of two fully synchronized clusters oscillating in antiphase with a phase difference of $\pi$ \cite{aoki2009co,aoki2011self}. The instantaneous phase snapshot in Fig.~\ref{AP}(a) clearly reveals the two antipodal clusters. The coupling matrix in Fig.~\ref{AP}(d) exhibits strong positive intra-cluster couplings and strong negative inter-cluster couplings, consistent with the Hebbian learning rule. The phase snap--shot, space--time plot and time-averaged frequency distribution shown in Figs.~\ref{AP}(a), \ref{AP}(b), and \ref{AP}(c) confirm the stability of this symmetric configuration.
\begin{figure}[!ht]
	\centering
	\includegraphics[width=0.5\textwidth]{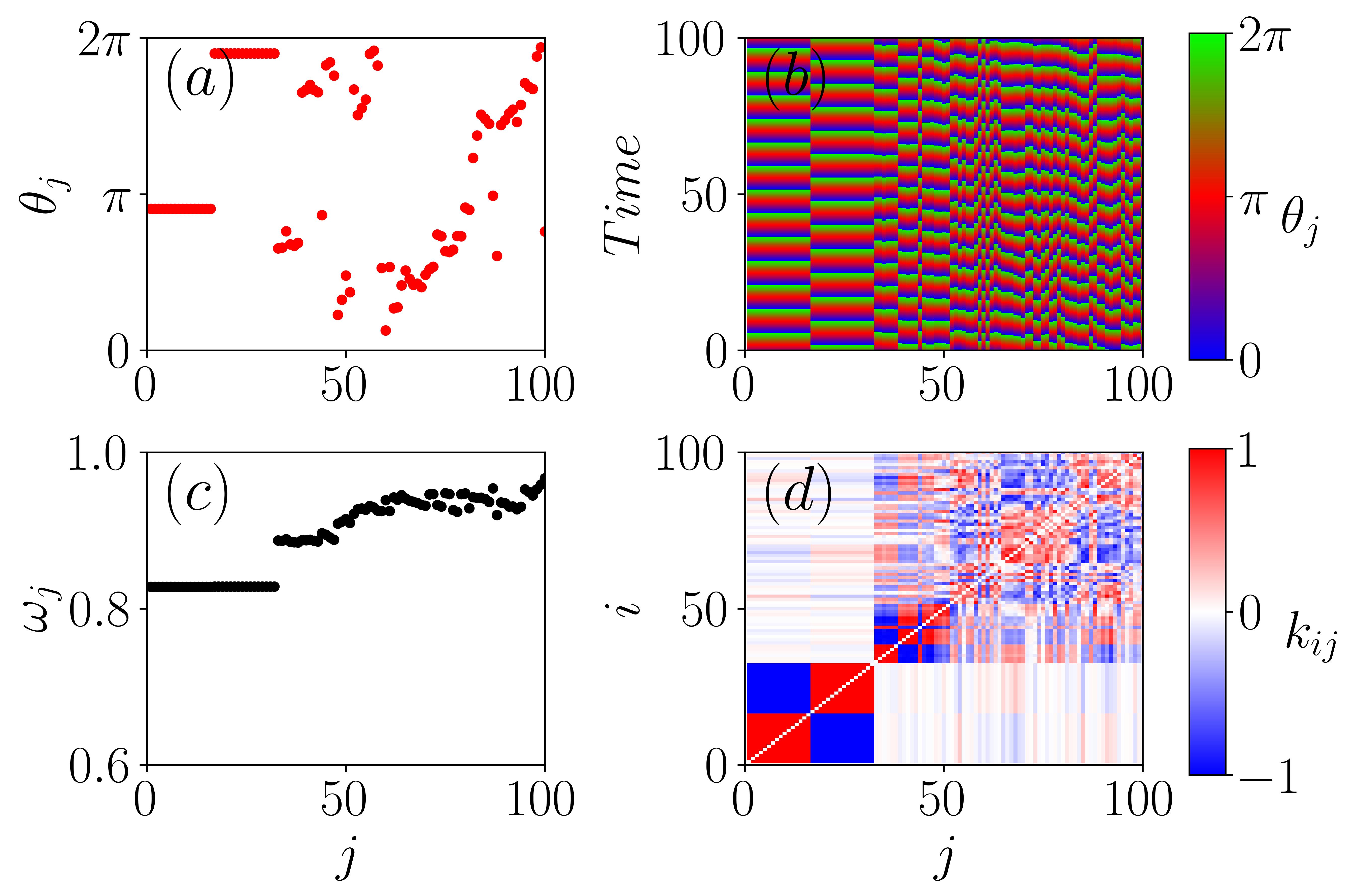}
	\caption{Chimera (CHI) state for $\alpha = 0.45\pi$ and $q = 0$: (a) snapshot of instantaneous phases, (b) space--time evolution of the oscillators, (c) time-averaged frequency distribution, and (d) coupling matrix $k_{ij}$. All other parameters are the same as in Fig.~\ref{FC1}.}
	\label{CHI}
\end{figure}
\begin{figure}[!ht]
	\centering
	\includegraphics[width=0.5\textwidth]{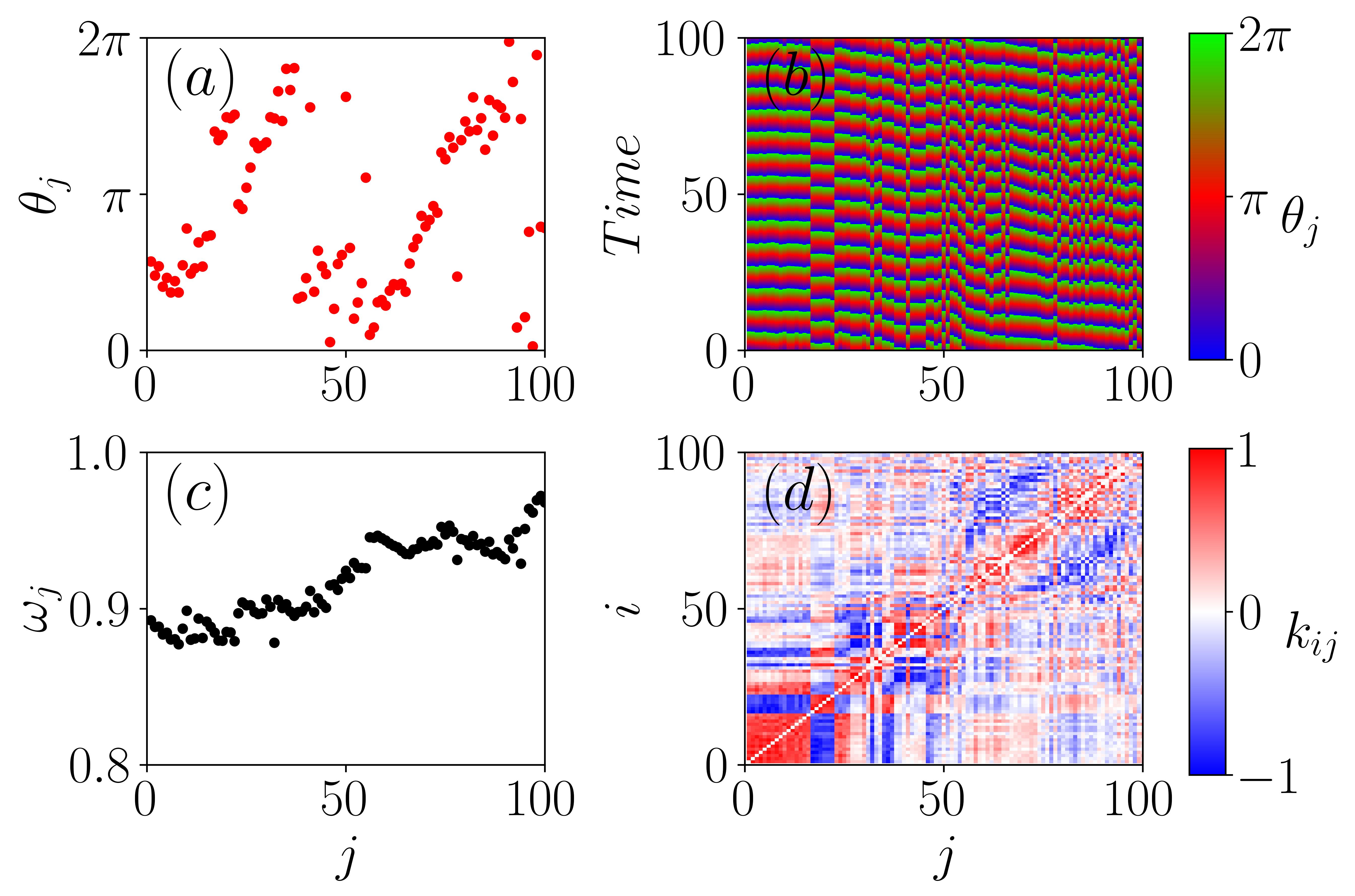}
	\caption{Incoherent (INC) state for $\alpha = 0.5\pi$ and $q = 0$: (a) snapshot of instantaneous phases, (b) space--time evolution of the oscillators, (c) time-averaged frequency distribution, and (d) coupling matrix $k_{ij}$. All other parameters are the same as in Fig.~\ref{FC1}.}
	\label{INC}
\end{figure}

Increasing the phase lag to $\alpha=0.25\pi$ leads to the emergence of a multi-antipodal cluster (MAC) state. In this regime, oscillators split into three or more clusters oscillating either in phase or in antiphase relative to one another. The instantaneous phase snapshot in Fig.~\ref{MAC}(a) reveals multiple evenly spaced clusters, while the space--time plot in Fig.~\ref{MAC}(b) confirms their long-term stability. The coupling matrix in Fig.~\ref{MAC}(d) displays alternating positive and negative blocks, reflecting excitatory intra-cluster and inhibitory inter-cluster interactions. The time-averaged frequency distribution in Fig.~\ref{MAC}(c) indicates that all clusters share nearly identical mean frequencies.
\begin{figure}[!ht]
	\centering
	\includegraphics[width=0.5\textwidth]{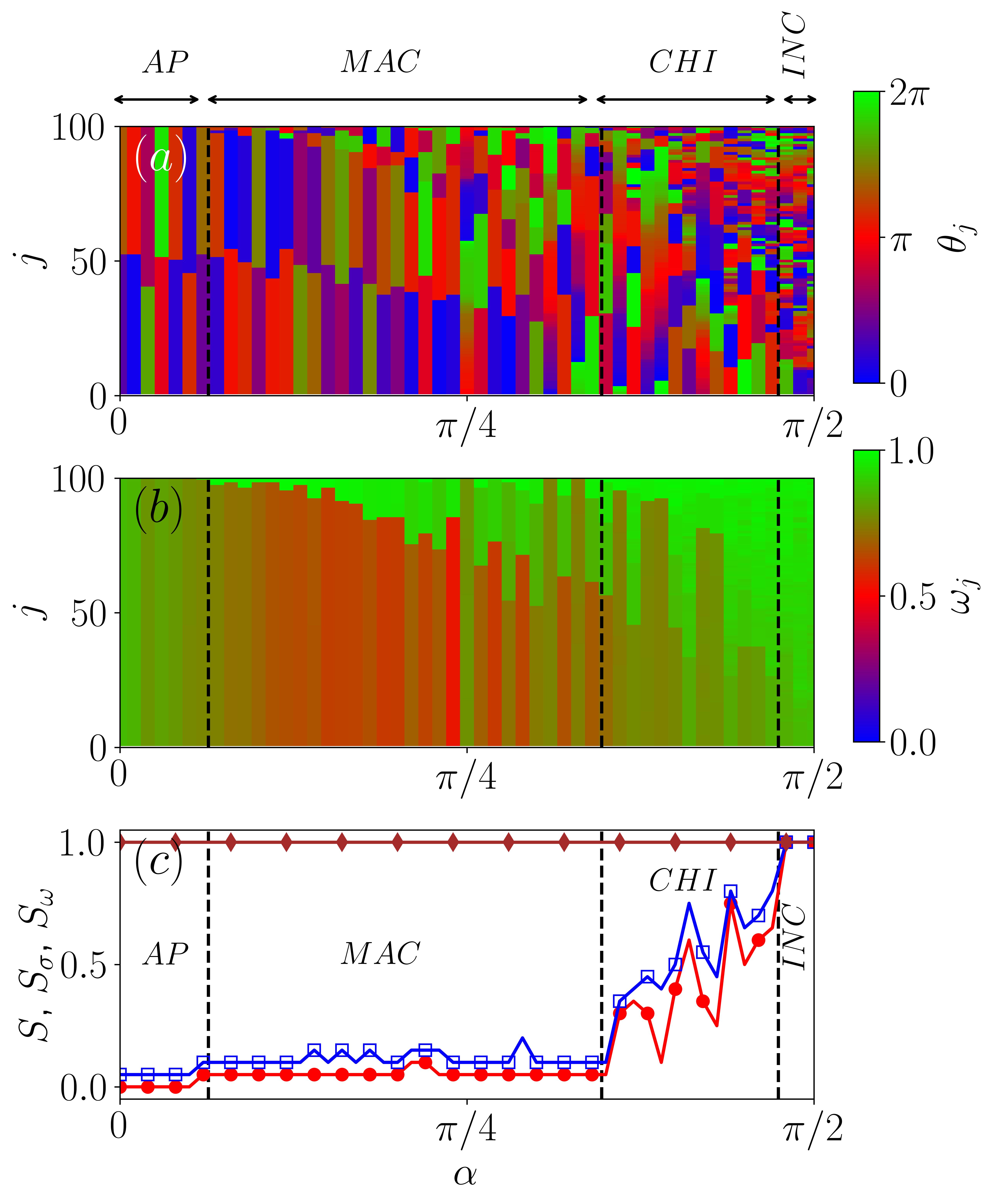}
	\caption{Dynamical transitions as a function of the phase lag $\alpha$ for $q = 0$: (a) instantaneous phases of the oscillators (color-coded by phase value), (b) time-averaged oscillator frequencies, and (c) variation of the three incoherence measures ($S$, $S_\sigma$, and $S_\omega$) with $\alpha$. All other parameters are the same as in Fig.~\ref{FC1}.}
	\label{1p2}
\end{figure}

For larger phase lag values ($\alpha=0.45\pi$), the system transitions into a chimera (CHI) state, characterized by the coexistence of synchronized and desynchronized subpopulations \cite{perc2021chimeras,majhi2019chimera}. The instantaneous phase snapshot, space--time plot, and time-averaged frequency distribution shown in Figs.~\ref{CHI}(a)--\ref{CHI}(c) clearly demonstrate this mixed coherent--incoherent organization. The corresponding coupling matrix in Fig.~\ref{CHI}(d) exhibits a partially ordered structure, reflecting the heterogeneous nature of the interactions.

For sufficiently large phase lag ($\alpha=0.5\pi$), the system loses all coherence and transitions into a fully incoherent (INC) state. In this regime, oscillators evolve independently, with randomly distributed phases and no collective synchronization. The disordered coupling matrix and the absence of structure in the phase snapshots, space--time plots, and frequency distributions are shown in Fig.~\ref{INC}.

The one-parameter diagram for the balanced PRC case is summarized in Fig.~\ref{1p2}. The AP state is observed for $\alpha\in(0,0.06\pi)$, followed by the MAC state for $\alpha\in(0.06\pi,0.35\pi)$ and the CHI state for $\alpha\in(0.35\pi,0.47\pi)$. Beyond $\alpha=0.47\pi$, the INC state dominates. These transitions are quantitatively supported by the incoherence measures shown in Fig.~\ref{1p2}(c).

\subsection{Positive PRC offset $(q>0)$}
\par
\begin{figure}[!ht]
	\centering
	\includegraphics[width=0.5\textwidth]{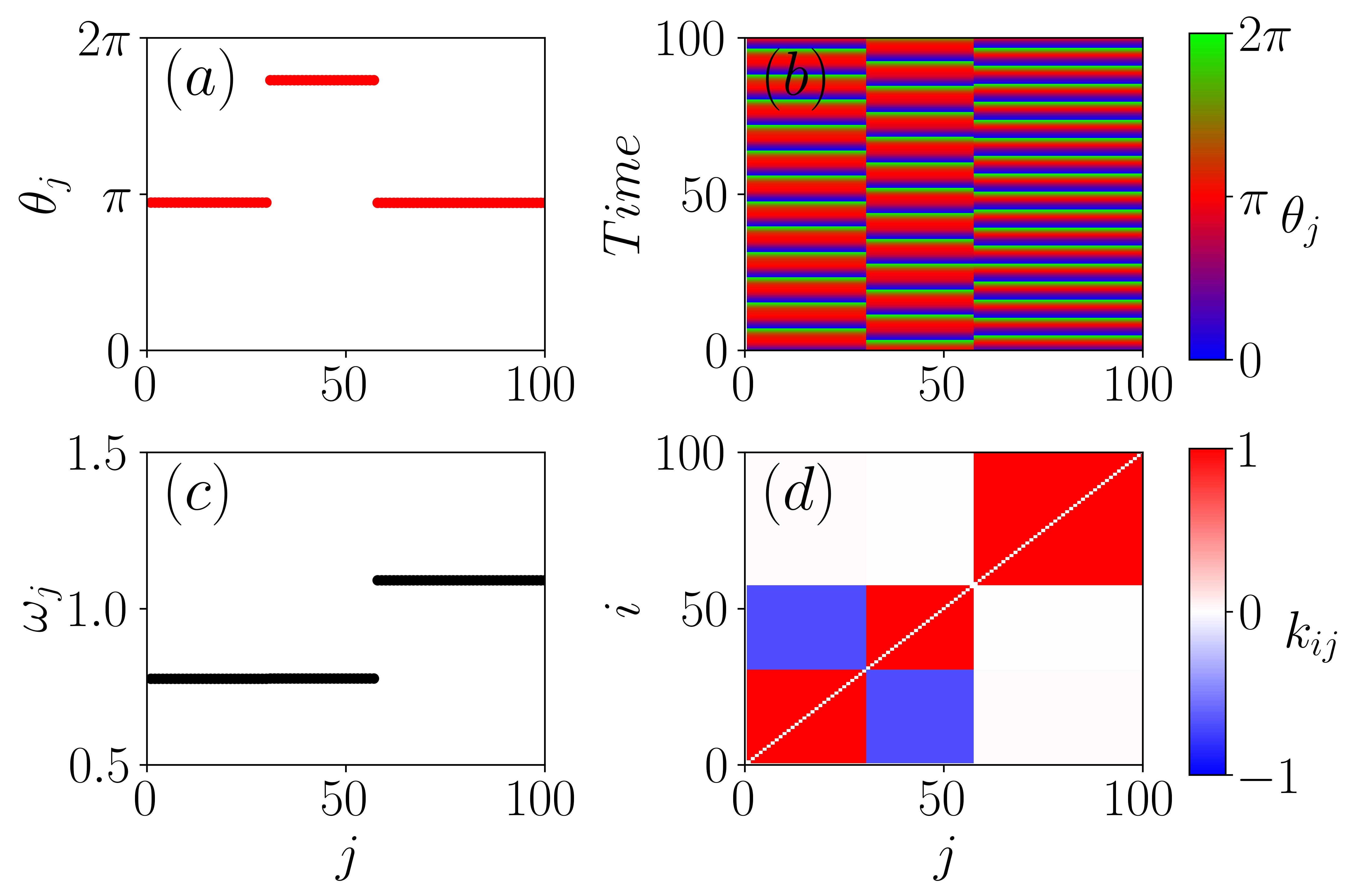}
	\caption{Frequency cluster (FC) state for $\alpha = 0$ and $q = 1$: (a) snapshot of instantaneous phases, (b) space--time evolution of the oscillators, (c) time-averaged frequency distribution, and (d) coupling matrix $k_{ij}$. All other parameters are the same as in Fig.~\ref{FC1}.}
	\label{FC2}
\end{figure}

 Finally, we consider the case of positive PRC offset ($q>0$). For $q=1$ and $\alpha=0$, the system again exhibits a frequency cluster (FC) state. The space--time plot in Fig.~\ref{FC2}(b) reveals persistent striped patterns corresponding to multiple frequency groups, while the instantaneous phase snapshot in Fig.~\ref{FC2}(a) and the frequency distribution in Fig.~\ref{FC2}(c) confirm the presence of frequency clustering. The corresponding coupling matrix is shown in Fig.~\ref{FC2}(d).

\begin{figure}[!ht]
	\centering
	\includegraphics[width=0.5\textwidth]{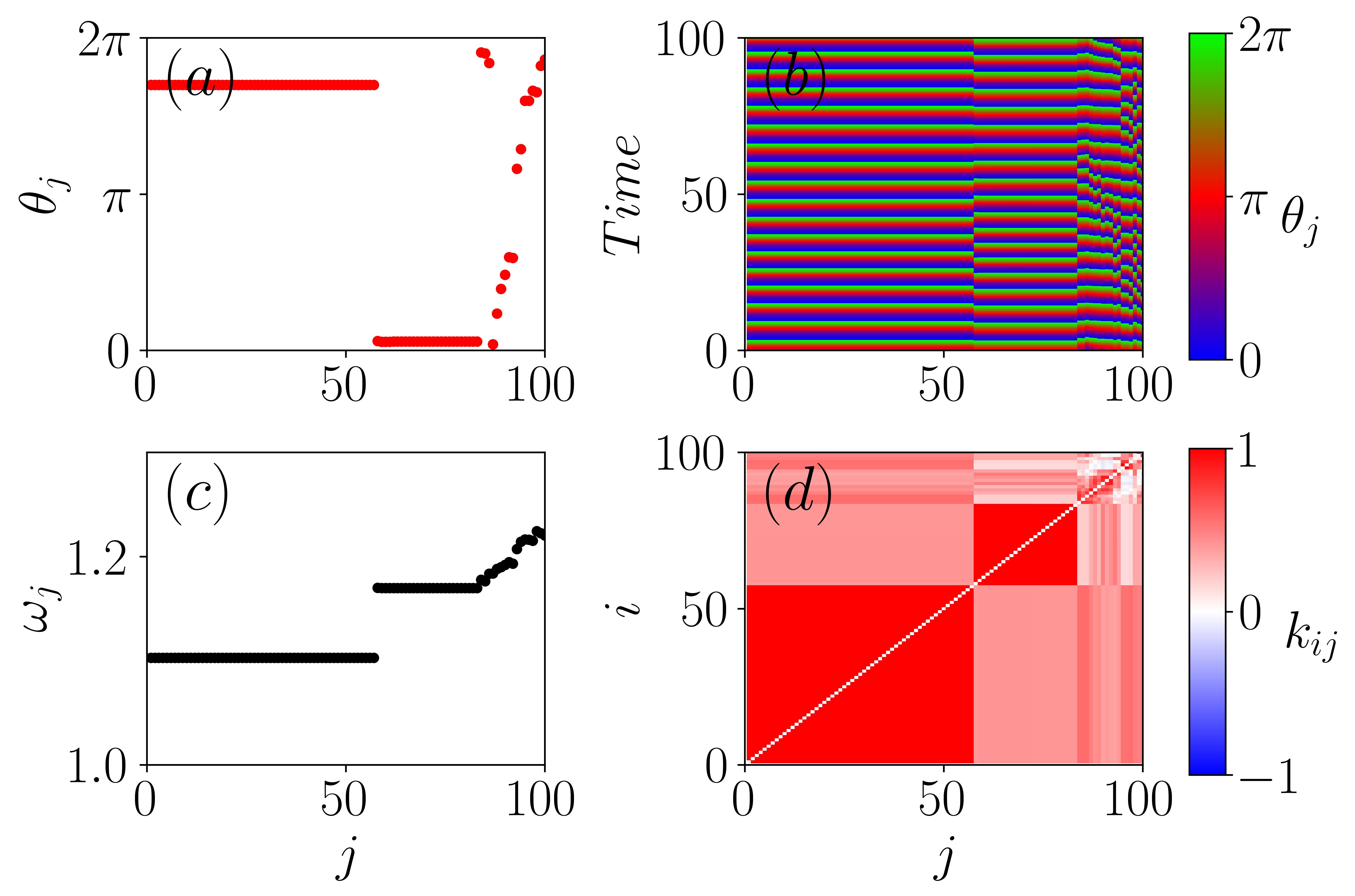}
	\caption{Chimera (CHI) state for $\alpha = 0.22\pi$ and $q = 1$: (a) snapshot of instantaneous phases, (b) space--time evolution of the oscillators, (c) time-averaged frequency distribution, and (d) coupling matrix $k_{ij}$.}
	\label{CHI2}
\end{figure}

Increasing the phase lag to $\alpha=0.22\pi$ induces a transition to the chimera (CHI) state. The coexistence of coherent and incoherent subpopulations is evident from the space--time plot in Fig.~\ref{CHI2}(b), the instantaneous phase snapshot in Fig.~\ref{CHI2}(a), and the heterogeneous coupling matrix in Fig.~\ref{CHI2}(d).

\begin{figure}[!ht]
	\centering
	\includegraphics[width=0.5\textwidth]{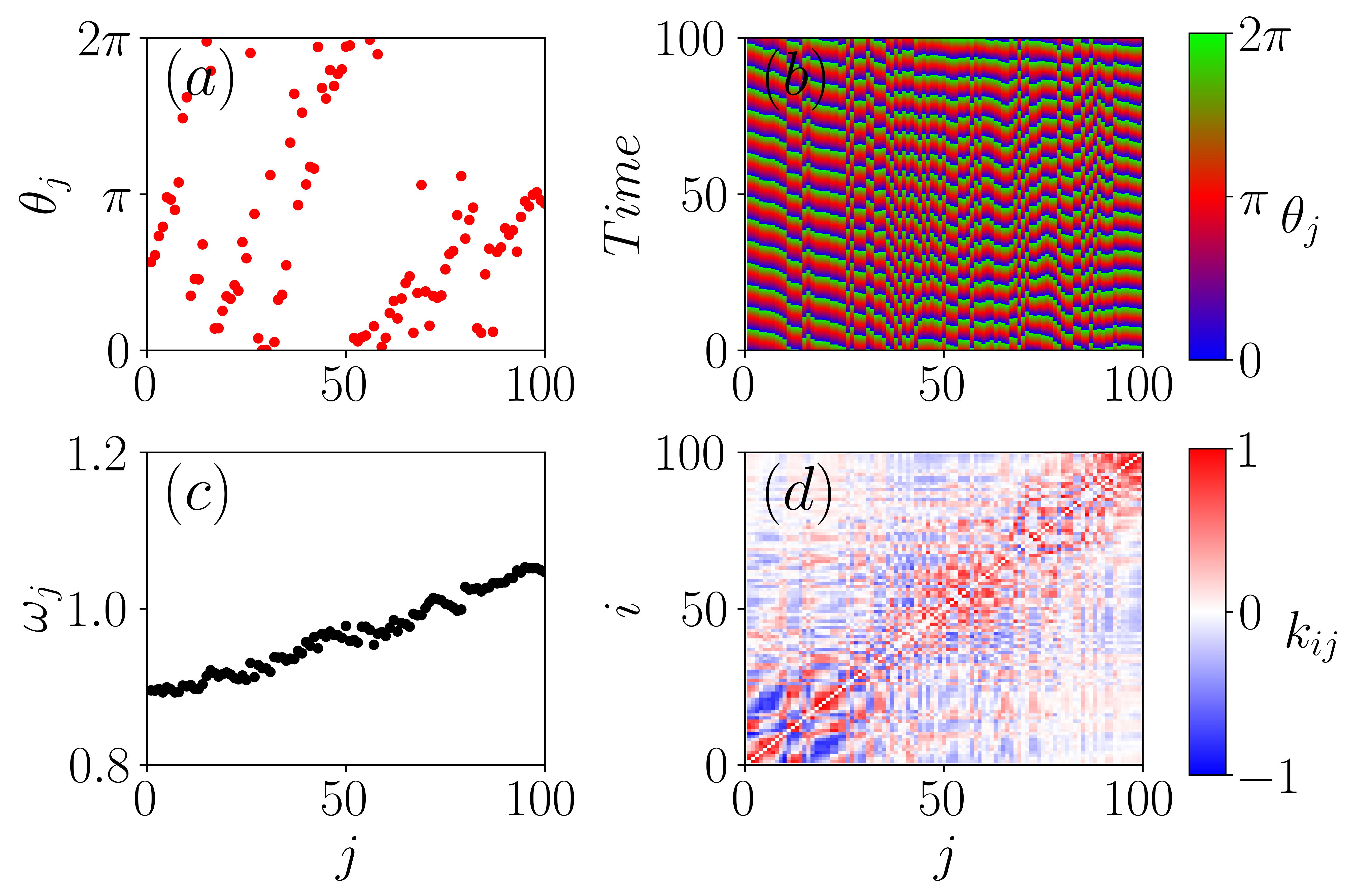}
	\caption{Incoherent (INC) state for $\alpha = 0.4\pi$ and $q = 1$: (a) snapshot of instantaneous phases, (b) space--time evolution of the oscillators, (c) time-averaged frequency distribution, and (d) coupling matrix $k_{ij}$.}
	\label{INC2}
\end{figure}

For larger phase lag values ($\alpha=0.4\pi$), the system transitions into a fully incoherent (INC) state, as shown in Fig.~\ref{INC2}. In this regime, the oscillators evolve independently, with random phases and disordered coupling strengths.

\begin{figure}[!ht]
	\centering
	\includegraphics[width=0.5\textwidth]{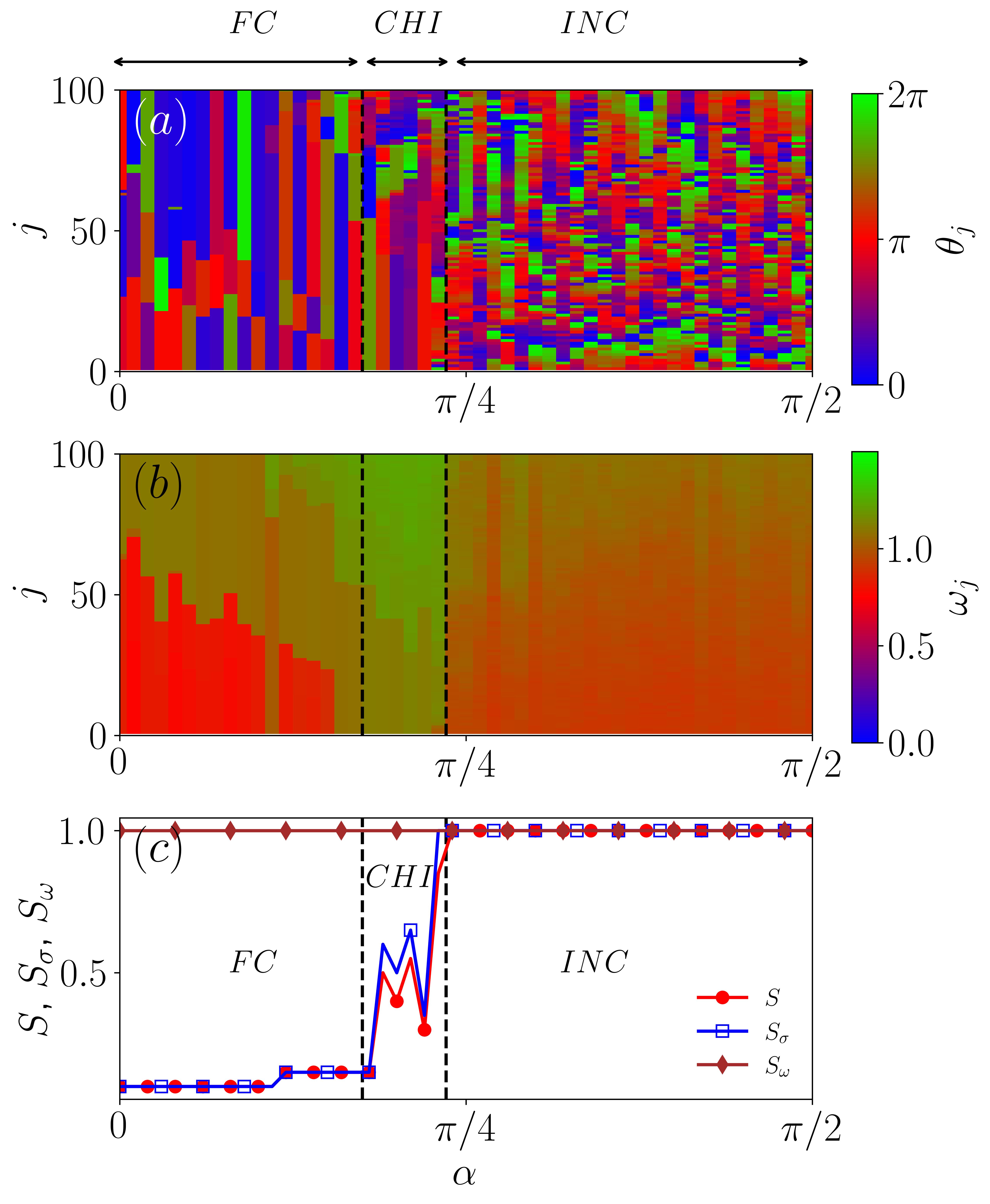}
	\caption{Dynamical transitions as a function of the phase lag $\alpha$ for $q = 1$: (a) instantaneous phases of the oscillators (color-coded by phase value), (b) time-averaged oscillator frequencies, and (c) variation of the three incoherence measures ($S$, $S_\sigma$, and $S_\omega$) with $\alpha$.}
	\label{1p3}
\end{figure}

The corresponding one-parameter diagram is shown in Fig.~\ref{1p3}. The FC state is observed for $\alpha\in(0,0.18\pi)$, followed by the CHI state for $\alpha\in(0.18\pi,0.23\pi)$. Beyond $\alpha=0.23\pi$, the system settles into the INC state. These transitions are corroborated by the incoherence measures shown in Fig.~\ref{1p3}(c), with $S_\omega=1$ throughout.
\begin{figure}[!ht]
	\centering
	\includegraphics[width=1.0\linewidth]{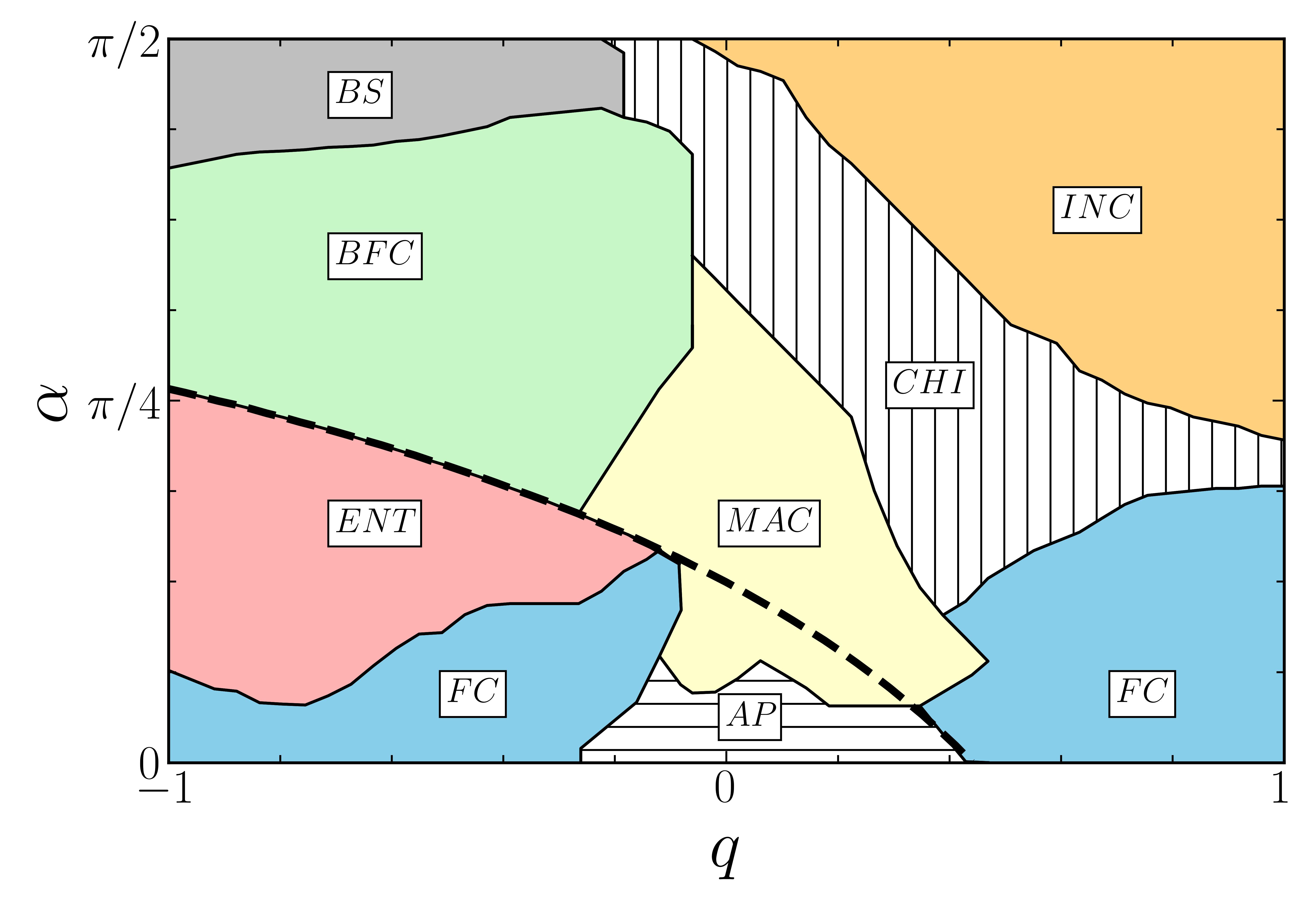}
	\caption{Two-parameter phase diagram illustrating the dynamical states in the $(q,\alpha)$ parameter space. The initial conditions for the phases $\theta_i$ and coupling strengths $k_{ij}$ are drawn uniformly from the intervals $[0,2\pi)$ and $(-1,1)$, respectively. The black dashed line denotes the analytical stability boundary for the frequency-entrained state.}
	\label{2p1}
\end{figure}

Finally, the global organization of the dynamical states across the $(q,\alpha)$ parameter plane is summarized in the two-parameter phase diagram shown in Fig.~\ref{2p1}. For negative PRC offsets $q\in(-1,-0.3)$, increasing $\alpha$ induces the sequence FC $\rightarrow$ ENT $\rightarrow$ BFC $\rightarrow$ BS, consistent with Fig.~\ref{1p1}. For intermediate offsets $q\in(-0.3,0.47)$, the system undergoes transitions AP $\rightarrow$ MAC $\rightarrow$ CHI, as shown in Fig.~\ref{1p2}. For positive offsets $q\in(0.47,1.0)$, the sequence FC $\rightarrow$ CHI $\rightarrow$ INC is observed, consistent with Fig.~\ref{1p3}. The black dashed line in Fig.~\ref{2p1} denotes the analytically obtained stability boundary $\alpha_c$ for the frequency-entrained state.	
\section{Analytical Approach} \label{Analytical}

In this section, we present an analytical treatment of the pulse-coupled adaptive network governed by Eqs.~\eqref{phase} and \eqref{coupling}, with the aim of classifying the observed dynamical states. While deriving analytical stability boundaries for all emergent states is not feasible due to the system’s complexity, the stability condition for the frequency-entrained state can be obtained explicitly.

Starting from Eqs.~\eqref{phase} and \eqref{coupling}, and fixing $\omega=1$ and $\sigma=1$, the governing equations reduce to
\begin{eqnarray}
	\label{winfree}
	\dot\theta_i &=& 1 + Q(\theta_i + \alpha)\frac{1}{N} \sum_{j = 1}^{N} k_{ij} P(\theta_j), \\
	\dot{k}_{ij} &=& \epsilon \bigl( \cos(\theta_i - \theta_j) - k_{ij} \bigr). \nonumber
\end{eqnarray}

In the frequency-entrained state, all oscillators lock to a common constant phase, $\theta_i=\theta$ for all $i$, corresponding to complete phase synchrony. Simultaneously, the adaptive coupling weights reach stationary values such that $\dot{k}_{ij}=0$ for all $i,j$. Under these conditions, Eq.~\eqref{winfree} simplifies to
\begin{eqnarray}
	\label{an1}
	1 + Q(\theta + \alpha)\frac{1}{N} \sum_{j = 1}^{N} k_{ij} P(\theta) &=& 0, \\
	\epsilon \bigl( \cos(0) - k_{ij} \bigr) &=& 0.
\end{eqnarray}

From the second equation, the fixed point of the coupling weights is obtained as
\begin{eqnarray}
	\label{an2}
	k^*_{ij} &=& 1.
\end{eqnarray}

The phase dynamics incorporates both the influence function $P(\theta)$ and the phase response curve $Q(\theta)$, as defined in Eqs.~\eqref{AS1} and \eqref{PRC1}. To determine stationary solutions, we seek fixed points where all oscillators share a common phase $\theta_i=\theta^*$ for all $i$. The corresponding reduced mean-field quantity $S$ becomes
\begin{eqnarray}
	\label{an4}
	S &=& 1 + \cos(\theta^*).
\end{eqnarray}

\begin{figure}[!ht]
	\centering
	\includegraphics[width=0.5\textwidth]{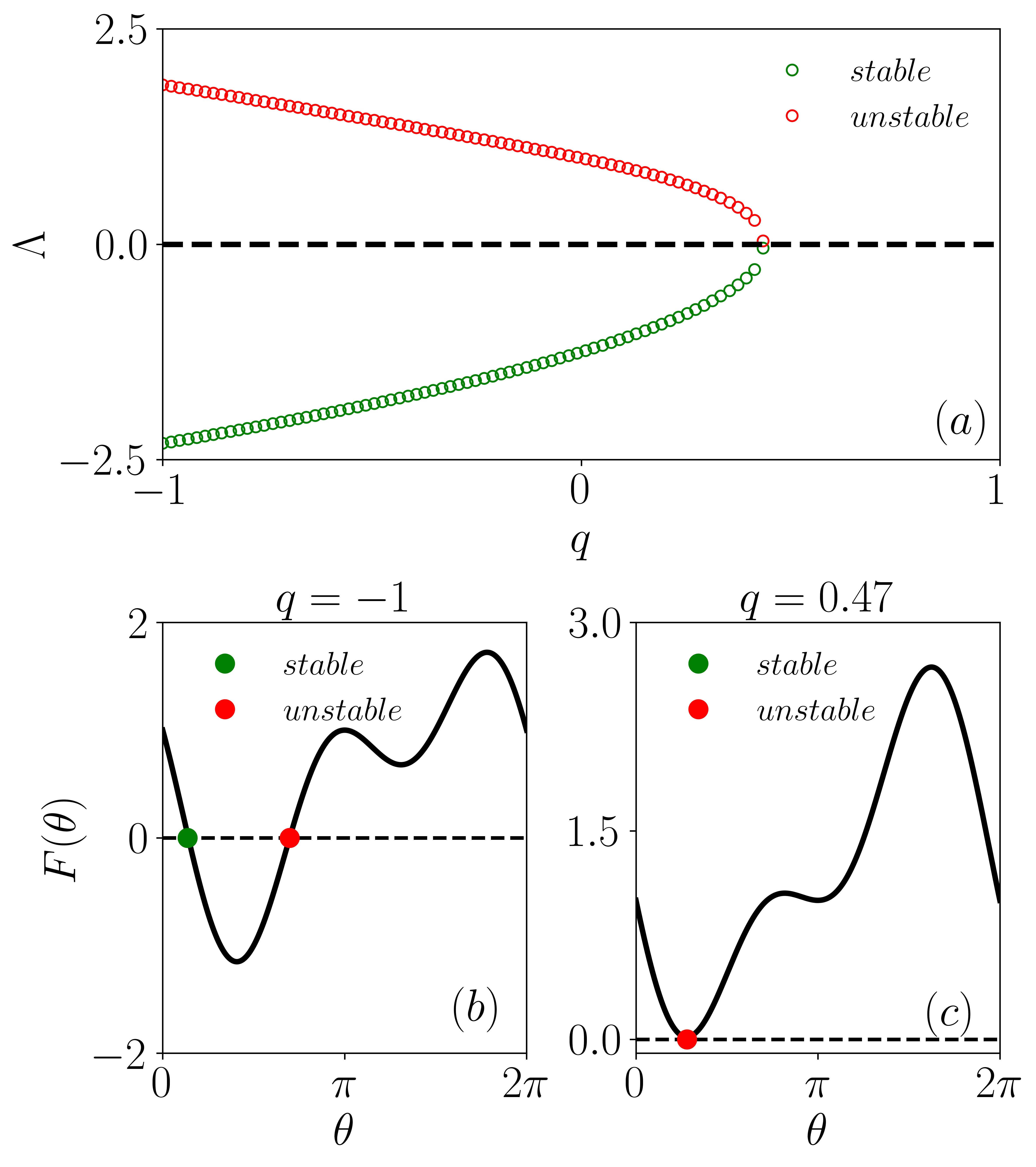}
	\caption{(a) Eigenvalues of the fixed points as a function of the PRC offset parameter $q$. Green open circles indicate stable fixed points, while red open circles indicate unstable ones. Panels (b) and (c) show the corresponding phase portraits for $q=-1$ and $q=0.47$, respectively, where stable and unstable fixed points are marked by green and red solid circles. The phase lag is fixed at $\alpha = 0.1\pi$.}
	\label{bif}
\end{figure}

Substituting the expressions for $P(\theta)$ and $Q(\theta)$ into Eq.~\eqref{an1}, the phase equilibrium condition can be written as
\begin{eqnarray}
	\label{an5}
	1 + \bigl[q(1 - \cos(\theta^* + \alpha)) - \sin(\theta^* + \alpha)\bigr](1 + \cos\theta^*) = 0. \nonumber
\end{eqnarray}

The stability of the fixed points $\theta^*$ and $k^*_{ij}$ is examined using linear stability analysis. To this end, we introduce small perturbations $\theta_i=\theta^*+\delta\theta_i$ and $k_{ij}=k^*_{ij}+\delta k_{ij}$. For notational convenience, we define
\[
A(\theta)=q(1-\cos(\theta+\alpha))-\sin(\theta+\alpha).
\]
The phase evolution equation then reads
\begin{eqnarray}
	\label{an6}
	\dot\theta_i = 1 + A(\theta_i)S_i.
\end{eqnarray}

Linearizing the coupling and phase-dependent terms yields
\begin{eqnarray}
	\label{an7}
	\delta \dot{k}_{ij} &=& -\epsilon \delta k_{ij}, \\
	A(\theta_i) &=& A(\theta^*) + A'(\theta^*)\delta\theta_i,
\end{eqnarray}
where $A'(\theta)=q\sin(\theta+\alpha)-\cos(\theta+\alpha)$. Since the perturbations $\delta k_{ij}$ decay exponentially, their contribution can be neglected at leading order. Retaining only first-order terms, the perturbation of the mean field becomes
\begin{eqnarray}
	\label{an8}
	\delta S_i = -\frac{\sin\theta^*}{N}\sum_{j=1}^{N}\delta\theta_j.
\end{eqnarray}

Consequently, the phase perturbations satisfy
\begin{eqnarray}
	\label{an9}
	\delta \dot{\theta}_i = S A'(\theta^*)\delta\theta_i - \frac{A(\theta^*)\sin(\theta^*)}{N}\sum_{j=1}^{N}\delta\theta_j.
\end{eqnarray}

Equation~\eqref{an9} has a diagonal-plus-rank-one structure, implying that the spectrum consists of $(N-1)$ degenerate transverse eigenvalues and one collective eigenvalue. For transverse perturbations satisfying $\sum_j \delta\theta_j=0$, the rank-one term vanishes, yielding
\begin{eqnarray}
	\label{an10}
	\mu_K = S A'(\theta^*).
\end{eqnarray}

For the collective (synchronous) perturbation mode, $\delta\theta_i=\delta\theta$ for all $i$, the eigenvalue is
\begin{eqnarray}
	\label{an11}
	\mu_N = S A'(\theta^*) - A(\theta^*)\sin(\theta^*).
\end{eqnarray}

Defining $F(\theta^*) = 1 + A(\theta^*)S$, it follows that
\begin{eqnarray}
	\label{an12}
	\mu_N = F'(\theta^*).
\end{eqnarray}

Frequency-entrained equilibria are created or annihilated through saddle-node (fold) bifurcations, at which the collective eigenvalue becomes marginal. This scenario is illustrated in Fig.~\ref{bif}. Figure~\ref{bif}(a) shows the eigenvalues of the fixed points as a function of the PRC offset parameter $q$, with stable and unstable fixed points indicated by green and red symbols, respectively. For $q=-1$ and $\alpha=0.1\pi$, two fixed points coexist, one stable and one unstable. As $q$ increases, these fixed points approach each other and collide at a saddle-node bifurcation near $q\simeq0.47$.

The fold condition is obtained from Eq.~\eqref{an11} as
\begin{eqnarray}
	\label{an13}
	F(\theta^*) = 0, \qquad F'(\theta^*) = 0.
\end{eqnarray}

Introducing the half-angle substitution $u=\tan(\theta/2)$ and $v=1+u^2\ge1$, we have
\begin{eqnarray}
	\label{an14}
	1 + \cos\theta = \frac{2}{v}, \qquad \sin\theta = \frac{2t}{v}.
\end{eqnarray}

Using these relations and the fold conditions in Eq.~\eqref{an13}, one finds $A(\theta)=-v/2$, which leads to the cubic equation
\begin{eqnarray}
	\label{an15}
	v^3 + 4qv - 4 = 0, \qquad v \ge 1.
\end{eqnarray}

For $q\in(-1,1)$, the physically relevant root $v(q)$ can be obtained numerically (e.g., using the Newton--Raphson method). The corresponding critical phase lag $\alpha_c(q)$ is then determined from the trigonometric relations implied by the fold conditions,
\begin{eqnarray}
	\label{an16}
	v \cos(\alpha_c) -(2-v)\sin(\theta + \alpha_c)+2u\cos(\theta + \alpha_c) = 0.
\end{eqnarray}

Equation~\eqref{an16} thus provides an implicit criterion for the critical phase lag $\alpha_c(q)$, which marks both the existence and stability boundary of the frequency-entrained state. This analytical boundary is shown as the black dashed line in the phase diagrams.
\begin{figure}[!ht]
	\centering
	\includegraphics[width=1.0\linewidth]{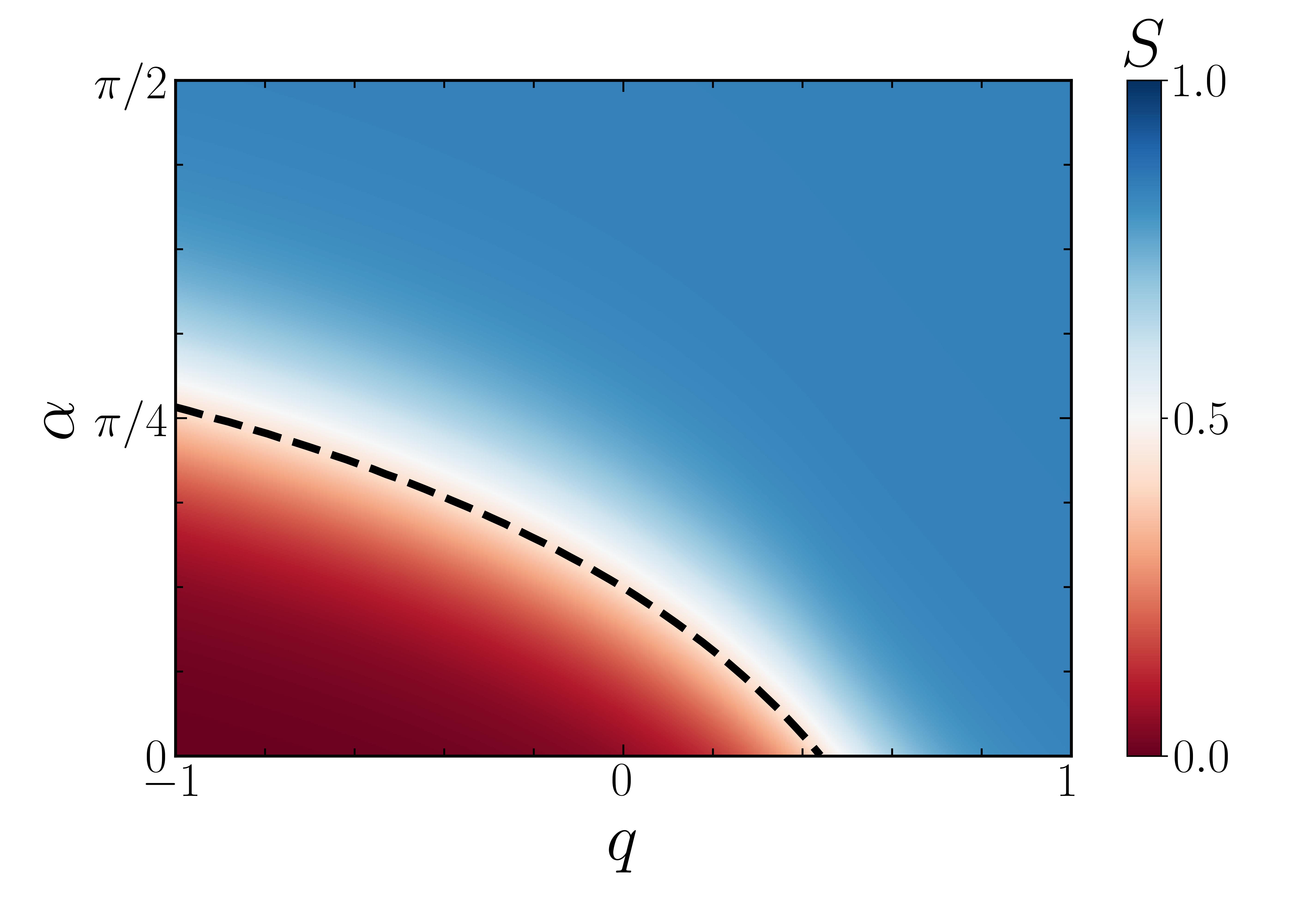}
	\caption{Two-parameter phase diagram in the $(q,\alpha)$ parameter space showing the strength of incoherence $S$, defined using the time-averaged frequency per bin. The black dashed line indicates the analytical stability boundary $\alpha_c$ of the frequency-entrained state.}
	\label{2p2}
\end{figure}

Figure~\ref{2p2} presents the numerically obtained strength of incoherence $S$ in the $(q,\alpha)$ parameter space. The frequency-entrained state corresponds to $S=0$, the fully incoherent state to $S=1$, and partially synchronized states to intermediate values $0<S<1$. Red regions indicate complete frequency entrainment, while blue regions denote full incoherence. The excellent agreement between numerical results and the analytical stability boundary $\alpha_c$ further validates the theoretical analysis.
	
\section{Conclusion}\label{conclusion}
In this work, we investigated a pulse-coupled Winfree neuron model endowed with a Hebbian adaptation rule. The adaptive coupling evolves in response to the global mean field, mediated through the phase response curve. The inclusion of a frustration (phase-lag) parameter $\alpha$ enables the emergence of a rich repertoire of self-organized dynamical patterns and their transitions.

The pulse-coupled adaptive network exhibits a wide variety of collective states, including frequency cluster, entrainment, bump-frequency cluster, bump, antipodal, multi-antipodal cluster, chimera, and incoherent states. These dynamical regimes were systematically characterized using three complementary measures of the strength of incoherence: (i) a frequency-based measure using time-averaged oscillator frequencies, (ii) a phase-based measure using instantaneous phases, and (iii) a mean-frequency-based measure evaluated per spatial bin.

The identification of these states was supported by multiple diagnostic tools, including space–time plots, coupling matrices, instantaneous phase snapshots, and time-averaged frequency profiles. Our results reveal intricate dynamical behavior arising from the interplay between the frustration parameter and the PRC offset. Specifically, negative PRC offsets promote synchronized and frequency-entrained patterns, whereas balanced and positive PRC offsets favor partially synchronized states and fully incoherent dynamics, respectively.

To systematically quantify these transitions, we employed the three incoherence measures $S$, $S_\sigma$, and $S_\omega$, which effectively distinguish between the observed dynamical regimes. A global overview of the state transitions was provided through two-parameter phase diagrams. For negative PRC offsets, the system undergoes transitions of the form FC $\rightarrow$ ENT $\rightarrow$ BFC $\rightarrow$ BS, consistent with the corresponding one-parameter bifurcation diagrams. For balanced PRC offsets, the observed transitions follow AP $\rightarrow$ MAC $\rightarrow$ CHI, while for positive PRC offsets the sequence FC $\rightarrow$ CHI $\rightarrow$ INC is observed, again in agreement with the one-parameter analyses.

A central and novel result of this study is the spontaneous emergence of entrainment, bump-frequency cluster, and bump states in an adaptive network governed solely by a Hebbian learning rule, without any external forcing. In contrast to earlier studies of adaptive Kuramoto-type models, where such states typically require external inputs, the pulse-coupled adaptive Winfree model supports these dynamical patterns through the intrinsic interplay between frustration and adaptive coupling. Furthermore, we analytically derived the stability criterion for the frequency-entrained state, and the resulting theoretical predictions are in excellent agreement with the numerically observed transition boundaries.

The emergence of diverse self-organized states, including entrainment, bump, and bump-frequency cluster states, in the absence of external forcing highlights the crucial role of Hebbian adaptation in pulse-coupled Winfree networks. The strong dependence of these states on the frustration parameter and PRC offset underscores how phase-lag-mediated plasticity can drive complex dynamical transitions. These findings emphasize the relevance of adaptive mechanisms for robust pattern formation and multistability in both biological and artificial networks \cite{markram2012spike,levina2007dynamical,diehl2015unsupervised}, with potential implications for understanding collective information processing in neuronal systems and for the design of neuromorphic computing architectures \cite{davies2018loihi,indiveri2011neuromorphic}.

Future research may extend this framework by incorporating intrinsic heterogeneity to investigate explosive synchronization and related transition phenomena. Additional directions include introducing phase lag simultaneously in the mean-field pulse and the PRC, as well as pursuing analytical studies in the continuum limit to uncover the mechanisms governing multistability and state formation. Identifying universal scaling laws associated with these transitions also remains an important open problem. Overall, this study provides deeper insight into how Hebbian adaptation and frustration jointly generate rich and complex dynamical behavior in pulse-coupled Winfree neuron models.

\section*{Acknowledgment}
R. A. acknowledges SASTRA for providing Teaching Assistantship. The research contributions of R. S. and V. K. C. are part of a project funded by the SERB-CRG (Grant No. CRG/2022/004784). The authors gratefully acknowledge the Department of Science and Technology (DST), New Delhi, for providing computational facilities through the DST-FIST program under project number SR/FST/PS-1/2020/135, awarded to the Department of Physics.

\section*{Data Availability}
The data that support the findings of this study are available from the corresponding author upon reasonable request.
\bibliography{ref}% Produces the bibliography via BibTeX.
\end{document}